\documentclass[12pt]{amsart}
\usepackage{amsthm}
\usepackage{amstext}
\usepackage{amssymb}
\usepackage{mathrsfs}
\usepackage{mathabx}
\usepackage[english]{babel}
\usepackage{enumerate}
\usepackage{xfrac}
\usepackage{mathabx}
\usepackage{mathtools}
\usepackage{xcolor}
\usepackage[all]{xy}

\usepackage[T1]{fontenc}

\theoremstyle{plain}
\newtheorem{theorem}{Theorem}[section]
\newtheorem{proposition}[theorem]{Proposition}
\newtheorem{corollary}[theorem]{Corollary}

\theoremstyle{definition}
\newtheorem{definition}[theorem]{Definition}

\newtheorem{remark}[theorem]{Remark}

\theoremstyle{remark}

\newcommand{\T}{\mathbb{T}}

\numberwithin{equation}{section}
\newcommand{\C}{\mathbb C}

\newcommand{\Id}{\mathrm{d}}

\newcommand{\IMM}{\mathscr{M}}

\newcommand{\IEE}{\mathscr{E}}

\newcommand{\Hom}{\operatorname{Hom}}

\newcommand{\End}{\operatorname{End}}

\newcommand{\dd}{\mathrm{d}}

\newcommand{\IC}{\mathbb{C}}
\newcommand{\IR}{\mathbb{R}}

\newcommand{\Cl}{\operatorname{\mathbb{C}l}}

\renewcommand\div{\mathrm{div}\thinspace}

\def\bf{\mathbf}

\newcommand{\ILL}{\mathscr{L}}

\newcommand{\IHH}{\mathscr{H}}

\newcommand{\IFF}{\mathscr{F}}
\newcommand{\IAA}{\mathscr{A}}

\newcommand{\ISS}{\mathscr{S}}

\newcommand{\dom}{\mathrm{Dom}}

\newcommand{\IN}{\mathbb{N}}
\newcommand{\IZ}{\mathbb{Z}}

\newcommand{\pa}{\slash\slash}

\newcommand{\bb}{\mathsf{b}}

\newcommand{\f}{\frac}

\newcommand\newdot{{\kern.8pt\cdot\kern.8pt}}
\def\nbull{{\raise1.5pt\hbox{\bf .}}}

\def\transport{/\hspace*{-3pt}/}

\title[Feynman-Kac formula for perturbations of order $\leq 1$]{Feynman-Kac formula for perturbations of order $\leq 1$, and noncommutative geometry}
\author{Sebastian Boldt}
\address{Mathematisches Institut\\Universit\"at Leipzig\\ Leipzig\\Germany}
\email{boldt@math.uni-leipzig.de}
\author{Batu G\"uneysu}
\address{Fakult\"at f\"ur Mathematik\\Technische Universit\"at Chemnitz\\Chemnitz\\Germany}
\email{batu.gueneysu@math.tu-chemnitz.de}

\begin{document}

\begin{abstract}
Let $Q$ be a differential operator of order $\leq 1$ on a complex metric vector bundle $\IEE\to \IMM$ with metric connection $\nabla$ over a possibly noncompact Riemannian manifold $\IMM$. Under very mild regularity assumptions on $Q$ that guarantee that $\nabla^{\dagger}\nabla/2+Q$ canonically induces a holomorphic semigroup $\mathrm{e}^{-zH^{\nabla}_{Q}}$ in $\Gamma_{L^2}(\IMM,\IEE)$ (where $z$ runs through a complex sector which contains $[0,\infty)$), we prove an explicit Feynman-Kac type formula for $\mathrm{e}^{-tH^{\nabla}_{Q}}$, $t>0$, generalizing the standard self-adjoint theory where $Q$ is a self-adjoint zeroth order operator. For compact $\IMM$'s we combine this formula with Berezin integration to derive a Feynman-Kac type formula for an operator trace of the form
$$
\mathrm{Tr}\left(\widetilde{V}\int^t_0\mathrm{e}^{-sH^{\nabla}_{V}}P\mathrm{e}^{-(t-s)H^{\nabla}_{V}}\Id s\right),
$$
where $V,\widetilde{V}$ are of zeroth order and $P$ is of order $\leq 1$. These formulae are then used to obtain a probabilistic representations of the lower order terms of the equivariant Chern character (a differential graded extension of the JLO-cocycle) of a compact even-dimensional Riemannian spin manifold, which in combination with cyclic homology play a crucial role in the context of the Duistermaat-Heckmann localization formula on the loop space of such a manifold. 
\end{abstract}

\subjclass[2020]{Primary 58J65, Secondary 47D03, 58J05}

\maketitle

%\tableofcontents

\section{Introduction}

The classical Feynman-Kac formula states that given a real-valued (for simplicity) smooth potential $V:\IMM\to\IR$ on a possibly noncompact Riemannian manifold $\IMM$ such that the symmetric Schrödinger operator $\Delta/2+V$ is semibounded from below in $L^2(\IMM)$ (defined initially on smooth compactly supported functions), one has 
$$
\mathrm{e}^{-t H_V}\Psi(x)= \mathbb{E}\left[1_{\{t<\zeta^x\}}\mathrm{e}^{-\int^t_0 V(\bb^x_s)\Id s} \Psi(\bb^x_t)\right]\quad\text{for all $\Psi\in L^2(\IMM)$, $t>0$, a.e. $x\in \IMM$,}
$$
whenever the expectation value is well-defined. Here
\begin{itemize}
\item $H_V$ denotes the Friedrichs realization\footnote{which corresponds to Dirichlet boundary conditions} of $\Delta/2+V$, taking into account that in general $\Delta/2+V$ need not have a unique self-adjoint realization, and $\mathrm{e}^{-t H_V}$ is defined via spectral calculus,
\item $\bb^x$ is an arbitrary Brownian motion on $\IMM$ starting from $x$ with lifetime $\zeta^x>0$, taking into account that $\IMM$ need not be stochastically complete.
\end{itemize}

Vector bundle versions of this formula have played a crucial role in mathematical physics through the Feynman-Kac-It\^{o} formula \cite{simon, bhl} and in geometry through probabilistic proofs of the Atiyah-Singer index theorem \cite{bismut, hsu}. In this context, one replaces $\Delta$ with $\nabla^{\dagger}\nabla$, where 
$$
\nabla:\Gamma_{C^{\infty}}(\IMM,\IEE)\longrightarrow \Gamma_{C^{\infty}}(\IMM,T^*\IMM\otimes\IEE)
$$
is a metric connection on a metric vector bundle $\IEE\to\IMM$, and the potential with a smooth pointwise self-adjoint section $V$ of $\mathrm{End}(\IEE)\to\IMM$. In other words, $V$ is a self-adjoint zeroth order operator. Assuming now that the symmetric covariant Schrödinger type operator $\nabla^{\dagger}\nabla/2+V$ in the space of square integrable sections $\Gamma_{L^2}(\IMM,\IEE)$ is bounded from below, one can prove that 
\begin{align}\label{asmmy}
\mathrm{e}^{-t H_V^{\nabla}}\Psi(x)= \mathbb{E}\left[1_{\{t<\zeta^x\}}\mathcal{V}^x_{\nabla}(t) \transport^x_{\nabla}(t)^{-1} \Psi(\bb^x_t)\right]\quad\text{for all $\Psi\in \Gamma_{L^2}(\IMM,\IEE)$, $t>0$, a.e. $x\in \IMM$,}
\end{align}
whenever the expectation is well-defined. Here
\begin{itemize}
\item $H_V^{\nabla}$ is the Friedrichs realization of $\nabla^{\dagger}\nabla/2+V$,
\item  $\transport^x_{\nabla}$ denotes the stochastic parallel transport along the paths of $\bb^x$ (cf. section \ref{hauss} below for the precise definition), 
\item  $\mathcal{V}^x_{\nabla}$ denotes the solution of the following pathwise given ordinary differential equation in $\mathrm{End}(\IEE_x)$,
$$
(\Id/\Id t)\mathcal{V}^x_{\nabla}(t)=- \mathcal{V}_{\nabla}^x(t)\transport^x_{\nabla}(t)^{-1}  V(\bb_t^x) \transport^x_{\nabla}(t),\quad \mathcal{V}_{\nabla}^x(0)=1.
$$
\end{itemize}
These facts are well-established (cf. the appendix of \cite{driver}). Note that a classical assumption on the negative part $V^-$ of $V$ that guarantees that $\nabla^{\dagger}\nabla/2+V$ is semibounded from below and that one has the uniform square-integrability 
\begin{align*}
\sup_{x\in \IMM} \mathbb{E}\left[1_{\{t<\zeta^x\}}|\mathcal{V}_{\nabla}^x(t)|^2\right]<\infty\quad\text{for all $t>0$}
\end{align*}
(so that by Cauchy-Schwarz the Feynman-Kac formula holds \cite{batuvil} for \emph{all} $f\in \Gamma_{L^2}(\IMM,\IEE)$) is given by $|V^-|\in\mathcal{K}(\IMM)$, the Kato class of $\IMM$ (cf. Definition \ref{QQQ}). Since bounded functions are always Kato, and since it is possible to find large (possibly weighted) $L^p+L^{\infty}$-type subspaces of $\mathcal{K}(\IMM)$ under very weak assumptions on the geometry of $\IMM$ (cf. Proposition \ref{PPP}), the Kato class becomes very convenient in the context of Feynman-Kac formulae and their applications.\vspace{2mm}

In contrast to the self-adjoint case, very little seems to be known concerning Feynman-Kac formulae in the situation where one replaces the self-adjoint zeroth order operator $V$ by an arbitrary differential operator $Q$ of order $\leq 1$, a situation that naturally leads to a non-self-adjoint theory. The aim of this paper is to provide a systematic treatement of this problem, dealing with all probabilistic and functional analytic problems that arise naturally in this context, mainly from the noncompactness of $\IMM$. Our essential insight here, which allows to detect the new probabilistic pieces of the Feynman-Kac formula \emph{explicitly} and which allows to deal with some of the functional analytic problems using perturbation theory, is to decompose $Q$ canonically in the form 
$$
Q= Q_{\nabla}+\sigma_1(Q)  \nabla,
$$
where 
$$
\sigma_1(Q)\in  \Gamma_{C^{\infty}}\big(\IMM,\Hom( T^{*}\IMM\otimes \IEE,\IEE)\big)
$$
denotes the first order principal symbol of $Q$, so that $Q_{\nabla}:=Q-\sigma_1(Q)$ is zeroth order. Since now $\nabla^{\dagger}\nabla+Q$ will typically not be symmetric in $\Gamma_{L^2}(\IMM,\IEE)$, we cannot use the Friedrichs construction to get a self-adjoint operator. Instead, we use Kato's theory of sectorial forms and operators (cf. appendix for the basics of sectorial forms/operators and holomorphic semigroups): to this end, we assume that $\nabla^{\dagger}\nabla/2+Q$ is \emph{sectorial}. It then follows from abstract results that this operator canonically induces a sectorial operator $H^{\nabla}_Q$ which generates a semigroup of bounded operators $\mathrm{e}^{-zH^\nabla_Q}$ in $\Gamma_{L^2}(\IMM,\IEE)$ which is holomorphic for $z$ running through some sector of the complex plane which contains $[0,\infty)$. For fixed $x\in\IMM$ let now $\mathcal{Q}^x_{\nabla}$ denote the solution to the It\^{o} equation
$$
\Id\mathcal{Q}^x_{\nabla}(t)=- \mathcal{Q}_{\nabla}^x(t)\transport^x_{\nabla}(t)^{-1} \big( \sigma_1(Q)^{\flat}(\Id\bb_t^x)+ Q_{\nabla}(\bb_t^x) \Id t\big)\transport^x_{\nabla}(t),\quad \mathcal{Q}_{\nabla}^x(0)=1,
$$
noting that one can give sense to the underlying It\^{o} differential $\sigma_1(Q)^{\flat}(\Id\bb_t^x)$ using the Levi-Civita connection on $\IMM$ (cf. Section \ref{hauss}). With these preparations, our main result, Theorem \ref{main} below, reads as follows: \vspace{2mm}

\emph{Let $\nabla^{\dagger}\nabla+Q$ be sectorial and let
\begin{align}\label{ass33}
\sup_{x\in K} \mathbb{E}\left[1_{\{t<\zeta^x\}}|\mathcal{Q}_{\nabla}^x(t)|^2\right]<\infty\quad\text{for all $K\subset\IMM$ compact, $t>0$.}
\end{align}
%is sectorial 
%$$
%\sigma_1(Q)\in \Gamma_{L^{\infty}}\big(\IMM,\Hom( T^{\dagger}\IMM, \mathrm{End}(\IEE))\big)
%$$
%and that one has the Kato property
%\begin{align}\label{ass3}
%\lim_{t\to 0+}\sup_{x\in\IMM}\int^T_0\mathbb{E}\left[1_{\{t<\zeta^x\}}|Q_{\nabla}(\bb^x_s)| \right]\Id s<1.
%\end{align}
Then for all $t>0$, $\Psi\in \Gamma_{L^2}(\IMM,\IEE)$, $x\in \IMM$, one has} 
\begin{align}\label{asscc}
\mathrm{e}^{-t H^{\nabla}_Q}\Psi(x) =\mathbb{E}\left[1_{\{t<\zeta^x\}}\mathcal{Q}_{\nabla}^x(t) \transport_{\nabla}^{x}(t)^{-1}\Psi(\mathsf{b}^x_t)\right].
\end{align}

Let us note that the locally uniform $L^2$-assumption (\ref{ass33}) serves two purposes: firstly, it decouples the validity of the Feynman-Kac formula from $\Psi$ (as in the above self-adjoint Kato situation). Secondly and more importantly, it allows us to conclude that the smooth representative of $\mathrm{e}^{-t H^{\nabla}_Q}\Psi$, which exists by local parabolic regularity, is in fact \emph{pointwise equal} to the right hand side of (\ref{asscc}), and not only almost everywhere. This is achieved by first proving the formula on relatively compact subsets of $\IMM$ using It\^{o}-calculus, and then letting these local formulae run through an exhaustion of $\IMM$, using a recent result for monotone convergence of nondensely defined sectorial forms (this procedure is, up to additional technical difficulties, somewhat analogous to the self-adjoint case) with a parabolic maximum principle for the heat equation (the use of which in this form being new even in the self-adjoint case). To the best of our knowledge, this pointwise identification of the smooth representative is new for stochastically incomplete $\IMM$'s even in the self-adjoint case. \vspace{2mm}

Making contact with perturbation theory through Kato type assumptions, in Proposition \ref{aaspopo} we prove:\vspace{2mm}

\emph{Assume either
\begin{itemize} 
\item $|\Re(\sigma_1(Q))|\in L^{\infty}(\IMM)$, 
\item $\Re(Q_{\nabla})$ is bounded from below by a constant $\kappa\in\IR$,
\item $|\Im(Q_{\nabla})|\in\mathcal{K}(\IMM)$,
\end{itemize}
or 
\begin{itemize} 
\item $\sigma_1(Q)$ is anti-selfadjoint and $|\sigma_1(Q)|\in L^{\infty}(\IMM)$,
\item $|\Re(Q_{\nabla})^-|\in \mathcal{K}(\IMM)$,
\item $|\Im(Q_{\nabla})|\in \mathcal{K}(\IMM)$.
\end{itemize}
Then $\nabla^{\dagger}\nabla+Q$ is sectorial, and one has
\begin{align}\label{appP}
\sup_{x\in \IMM} \mathbb{E}\left[1_{\{t<\zeta^x\}}|\mathcal{Q}_{\nabla}^x(t)|^2\right]<\infty\quad\text{for all $t>0$}.
\end{align}
In particular, (\ref{asscc}) holds true.}\vspace{2mm}

Note that above $\Re(A)$ and $\Im(A)$ denote, respectively, the fiberwise defined real part and imaginary part of any zeroth order operator. Since these are self-adjoint zeroth order operators, one can define their positive/negative parts using the spectral calculus fiberwise. Note that, while in the self-adjoint case one can control $|\mathcal{Q}^x_{\nabla}(t)|$ pathwise using Gronwall's inequality, in the situation of Theorem \ref{main} and  Proposition \ref{aaspopo} one has to estimate the solution of a covariant It\^{o}-equation, which in combination with the noncompactness of $\IMM$ leads to several technical difficulties. Although the present formulation of Proposition \ref{aaspopo} should cover most applications, it would be natural to replace any (lower) boundedness assumption in Proposition \ref{aaspopo} with an appropriate Kato-type assumption. Although we tried hard, we have not been able to do that. It would also be very interesting to obtain non self-adjoint variants of semigroup domination \cite{berard,driver,ouhabaz,shigekawa} (also called 'Kato-Simon inequality' in \cite{batu2}) using the Feynman-Kac formula in the above setting, keeping in mind that such estimates play a crucial role in geometric analysis (see e.g. \cite{pallara,boldt}) and in mathematical physics (where they are called 'diamagnetic inequalities' \cite{simon1,bhl}). In the self-adjoint case these estimates take the form 
$$
|\mathrm{e}^{-t H^{\nabla}_V}\Psi(x)|\leq \mathrm{e}^{-t H_v}|\Psi|(x),
$$
where $v:\IMM\to\IR$ is any scalar potential such that for all $x\in \IMM$ every eigenvalue of $V(x)$ is $\geq v(x)$.\\
It should also be noted that, if one ignores functional analytic problems that arise for example from the noncompactness of $\IMM$, it is somewhat natural that \emph{some} probabilistic representation of $\mathrm{e}^{-t H^{\nabla}_Q}$ must exist: as $\nabla$ is metric, the operator $\nabla^{\dagger}\nabla+Q$ equals $-\mathrm{tr}\nabla^2+Q$, and (see appendix, Section \ref{bochner}), the latter nondivergence form operator can be canonically rewritten in the nondivergence form $-\mathrm{tr}\widetilde{\nabla}^2+\widetilde{Q}$, where $\widetilde{\nabla}$ is another connection and $\widetilde{Q}$ is of \emph{zeroth} order (keeping in mind that at least $\widetilde{Q}$ is somewhat implicitly given; see also Proposition 2.5 in \cite{getzler}). For compact $\IMM$'s no particular analytic problems arise, and the Feynman-Kac formula for $-\mathrm{tr}\widetilde{\nabla}^2+\widetilde{Q}$ is formally of the type (\ref{asmmy}), as shown in \cite{norris} (section 8 therein). On the other hand, in our noncompact setting, the divergence form $\nabla^{\dagger}\nabla+Q$ is favourable from an analytic point of view, and (see again appendix, Section \ref{bochner}) in this case it is in general not possible to rewrite this operator in the divergence form $\widetilde{\nabla}^\dagger \widetilde{\nabla} +\widetilde{Q}$, with $\widetilde{Q}$ zeroth order. From this point of view, we believe that our formulation of the Feynman-Kac formula is optimal in the noncompact case from an analytic point of view. Moreover, our formula has even some advantages in some applications to compact $\IMM$'s, where the generator appears precisely in the form $\nabla^\dagger\nabla +\sigma_1(Q)  \nabla+Q_{\nabla}$ (see below).\vspace{2mm}

Our next main result is the following trace formula (cf. Theorem \ref{main2}):\vspace{2mm}

\emph{Assume $\IMM$ is compact, and let $P$ be of order $\leq 1$, and let $V,\widetilde{V}$ be of zeroth order. Then for all $t>0$ one has\small{
\begin{align}\label{pqmxA}
&\mathrm{Tr}\left(\widetilde{V}\int^t_0\mathrm{e}^{-sH^{\nabla}_{V}}P \mathrm{e}^{-(t-s)H^{\nabla}_{V}}\Id s \right)\\\nonumber
&=-\int_{\IMM}\mathrm{e}^{-tH}(x,x)\mathrm{Tr}_{x}\left( \widetilde{V}(x)\mathbb{E}^{x,x}_t\left[\mathcal{V}^x_{\nabla}(t)\int^t_0 \transport^x_{\nabla}(s)^{-1} \big( \sigma_1(P)^{\flat}(\Id\bb_s^x)+ P_{\nabla}(\bb_s^x) \Id s\big)\transport^x_{\nabla}(s)\pa^{x}_{\nabla}(t)^{-1}\right]\right) \Id\mu(x),
\end{align}}
where $\mathrm{e}^{-tH}(x,y)$ denotes the integral kernel of the Friedrichs realization of $\Delta$ (in other words, the heat kernel on $\IMM$), and $\mathbb{E}^{x,x}_t$ denotes the expectation with respect to the Brownian bridge starting in $x$ and ending in $x$ at the time $t$. }\vspace{2mm}

The proof of this result is in fact reduced to (\ref{asscc}) using Berezin integration, a trick which has been communicated to the authors by Shu Shen. It would be very interesting to see, if at least for certain $P$'s it is possible to obtain (\ref{pqmxA}) using the very general Bismut derivative formulae from \cite{driver} in combination with the Markov property of Brownian motion. We have not worked into this direction.\vspace{2mm}

Finally, we use (\ref{pqmxA}) together with a new commutation formula for spin-Dirac operators (cf. formula (\ref{ssppaa}) below) to establish a probabilistic formula for the 'first order' part of the equivariant Chern-Character $\mathrm{Ch}_{\mathbb{T}}(\IMM)$ of a compact even-dimensional Riemannian spin manifold $\IMM$, where $\mathbb{T}:=S^1$. We refer the reader to Section \ref{aaQQ} for the definition of $\mathrm{Ch}_{\mathbb{T}}(\IMM)$ and concentrate here only the probabilistic side of the formula: to this end,  note that every element $\alpha$ of the space $\Omega_\T(\IMM)$ of $\mathbb{T}$-invariant differential forms on $\IMM\times \mathbb{T}$ can be uniquely written in the form $\alpha=\alpha'+\alpha''\Id t$ with $\Id t$ the volume form on $\T$. Then $\mathrm{Ch}_{\mathbb{T}}(\IMM)$ becomes a complex linear functional on the space
%$$
% \mathsf{C}_\T(\IMM): = \bigoplus_{N=0}^\infty\Omega_\T(\IMM)^{\otimes (N+1)}.
%$$
$$
\mathsf{C}_\T(\IMM): = \bigoplus_{N=0}^\infty\Omega_\T(\IMM)\otimes\left(\Omega_\T(\IMM)^{\otimes N} /  (\C\cdot 1)  \right).
$$

In Theorem \ref{main3} we prove:\vspace{2mm}

\emph{For all $\alpha_0,\alpha_1\in \Omega_\T(\IMM)$, $t>0$ one has
\begin{tiny}
\begin{align*}
&\mathrm{Ch}_{\mathbb{T}}(\IMM)(\alpha_0\otimes\alpha_1)\\
&=\int_\IMM\mathrm{e}^{-tH}(x,x)\mathrm{Str}_x\left(c(\alpha_0')(x)\mathbb{E}^{x,x}_t\left[\mathrm{e}^{-(1/8)\int^t_0\mathrm{scal}(\bb^x_s)\Id s}\int^t_0 \transport^x_{\nabla}(s)^{-1} \Big( 2c(*\Id\bb_s^x\lrcorner \alpha_1')  -c(\alpha_1'')(\bb_s^x) \Id s\Big)\transport^x_{\nabla}(s)\transport^{x}_{\nabla}(t)^{-1}\right]|_{t=2}\right)\Id\mu(x),
\end{align*}
\end{tiny}
where
\begin{itemize}
\item $\mathrm{Str}_x$ denotes the $\mathbb{Z}_2$-graded trace on $\mathrm{End}(\ISS_x)$, with $\ISS\to\IMM$ the spin bundle,
\item $\transport^x_{\nabla}$ denotes the stochastic parallel transport $\ISS\to\IMM$,
\item $c:\Omega_{C^{\infty}}(\IMM)\to \Gamma_{C^{\infty}}(\IMM,\mathrm{End}(\ISS))$ denotes Clifford multiplication,
\item $c(*\Id\bb_s^x\lrcorner \alpha)$ denotes a \emph{Stratonovic} differential with respect to the $\mathrm{End}(\ISS)$-valued $1$-form $v\mapsto c(v\lrcorner \alpha)$,  
\item $\mathbb{E}^{x,x}_t$ denotes the expectation with respect to the Brownian bridge starting $x$ and ending at the time $t$ in $x$.
\end{itemize}
}

We remark that $\mathrm{Ch}_{\mathbb{T}}(\IMM)$ has been introduced in \cite{gl} in the abstract setting of $\vartheta$-summable Fredholm modules over locally convex differential graded algebras and is in fact a differential-graded refinement of the JLO-cocycle \cite{jlo} for ungraded algebras. When applied to a compact even dimensional Riemannian spin-manifold, this construction provides via Chen integrals an algebraic model for Duistermaat-Heckman localization on the space of smooth loops, allowing a proof of the Atiyah-Singer index theorem for twisted spin-Dirac operators in the spirit of Atiyah \cite{atiyah} and Bismut \cite{bismut2}. We refer the reader to the introduction of \cite{gl} for a detailed explanation of these results. Obtaining a probabilistic formula for the higher order pieces of the equivarant Chern character remains an open problem at this point. \vspace{3mm}

\emph{Acknowledgements:} The authors would like to thank Shu Shen for a very helpful discussion that lead to the proof of Theorem \ref{main2}. Furthermore, we are grateful to the anonymous referee for several remarks that helped to improve the presentation of the paper.

\section{Main results}\label{hauss}

Let $\IMM$ be a connected Riemannian manifold of dimension $m$, where we work exclusively in the catogory of smooth manifolds without boundary. As such it is equipped with its Levi-Civita connection and its volume measure $\mu$. We denote the open geodesic balls with $B(x,r)\subset \IMM$.\\
Any fiberwise metric on a vector bundle will simply be denoted with $(\bullet,\bullet)$, with $|\bullet|:=\sqrt{(\bullet,\bullet)}$.\\
If $\IEE\to \IMM$ is a metric vector bundle and $p\in [1,\infty]$, then the norm on the complex Banach space of $L^p$-sections is denoted with
$$
\left\|\Psi\right\|_p:=\left(\int |\Psi|^p\Id\mu\right)^{1/p}.
$$
(with the obvious replacement for $p=\infty$). The scalar product in the Hilbert space $\Gamma_{L^2}(\IMM,\IEE)$ is denoted by 
$$
\left\langle \Psi_1,\Psi_2\right\rangle=\int (\Psi_1,\Psi_2) \Id \mu.
$$
Given another metric vector bundle $\IFF\to \IMM$ and a differential operator 
$$
P:\Gamma_{C^{\infty}}(\IMM,\IEE)\longrightarrow \Gamma_{C^{\infty}}(\IMM,\IFF)
$$
of order $\leq k$ with smooth coefficients, its formal adjoint 
$$
P^{\dagger}:\Gamma_{C^{\infty}}(\IMM,\IEE)\longrightarrow \Gamma_{C^{\infty}}(\IMM,\IFF)
$$
is the uniquely determined differential operator of order $\leq k$ with smooth coefficients, which satisfies
$$
\left\langle P\Psi_1,\Psi_2\right\rangle=\left\langle \Psi_1,P^{\dagger}\Psi_2\right\rangle\quad\text{for all}\quad\Psi_1\in \Gamma_{C^{\infty}_c}(\IMM,\IEE), \Psi_2\in \Gamma_{C^{\infty}_c}(\IMM,\IEE).
$$

Assume from now on that $\IEE\to \IMM$ is a metric vector bundle with a smooth metric connection 
$$
\nabla:\Gamma_{C^{\infty}}(\IMM,\IEE)\longrightarrow \Gamma_{C^{\infty}}(\IMM,T^*\IMM\otimes\IEE)
$$
Given a differential operator
$$
Q:\Gamma_{C^{\infty}}(\IMM,\IEE)\longrightarrow \Gamma_{C^{\infty}}(\IMM,\IEE)
$$
of order $\leq 1$, then with its first order principal symbol
$$
\sigma_1(Q)\in \Gamma_{C^\infty}\big(\IMM,\Hom( T^{*}\IMM, \mathrm{End}(\IEE))\big)=  \Gamma_{C^\infty}\big(\IMM,\Hom( T^{*}\IMM\otimes \IEE,\IEE)\big),
$$
the operator 
$$
Q_{\nabla}:=Q-\sigma_1(Q)\nabla\quad\text{is zeroth order,} 
$$
thus 
$$ 
Q_{\nabla}\in \Gamma_{C^\infty}(\IMM,\mathrm{End}(\IEE)),\quad Q= Q_{\nabla}+\sigma_1(Q)  \nabla.
$$
Assume that for every $x\in \IMM$ we are given a maximally defined Brownian motion 
$$
\bb^x:[0,\zeta^x)\times\Omega\longrightarrow\IMM
$$
on $\IMM$ with starting point $x$ and explosion time $\zeta^x>0$, which is defined on a fixed filtered probability space $(\Omega,\IFF,\IFF_*,\mathbb{P})$ that satisfies the usual assumptions. Let 
$$
\transport^x_{\nabla}:[0,\zeta^x)\times\Omega\longrightarrow  \IEE\boxtimes \IEE^{\dagger}
$$
be the corresponding stochastic parallel transport with respect to the fixed metric connection, where $\IEE\boxtimes \IEE^{\dagger}\to\IMM\times\IMM$ denotes the vector bundle whose fiber at $(a,b)$ is $\Hom(\IEE_a,\IEE_b)$. This is the uniquely determined continuous semimartingale such that \cite{norris} for all $t\in [0,\zeta^x)$,
\begin{itemize}
\item one has $\transport_{\nabla}^x(t):\IEE_x\to\IEE_{\bb_t(x)}$ unitarily, 
\item for all $\Psi\in \Gamma_{C^{\infty}}(\IMM,\IEE)$ one has
\begin{align}\label{para}
\transport_{\nabla}^{x}(t)^{-1}\Psi(\bb_t^x)= \transport_{\nabla}^{x}(t)^{-1}\nabla(*\Id\bb_t^x)\Psi(\bb_t^x),\quad \transport_{\nabla}^x(0)=1.
\end{align}
\end{itemize}
Above and in the sequel, $*\Id $ stands for Stratonovic integration, while $\Id$ will denote It\^{o} integration. Note that one can integrate $1$-forms in the Stratonovic sense on any manifold along any continuous semimartingale, while one can integrate $1$-forms on $\IMM$ along $\bb^x$ also in the It\^{o} sense, using the Levi-Civita connection on $\IMM$.\\
Define the process
$$
\mathcal{Q}^x_{\nabla}:[0,\zeta^x)\times\Omega\longrightarrow\mathrm{End}(\IEE_x)
$$
as the unique solution to the It\^{o} equation
$$
\Id\mathcal{Q}^x_{\nabla}(t)=- \mathcal{Q}_{\nabla}^x(t)\transport^x_{\nabla}(t)^{-1} \big( \sigma_1(Q)^{\flat}(\Id\bb_t^x)+ Q_{\nabla}(\bb_t^x) \Id t\big)\transport^x_{\nabla}(t),\quad \mathcal{Q}_{\nabla}^x(0)=1.
$$
Written out explicitly, the above equation means that for all $t\geq 0$ one has  
$$
\mathcal{Q}^x_{\nabla}(t)=1- \int^t_0\mathcal{Q}_{\nabla}^x(s)\transport^x_{\nabla}(s)^{-1}  \sigma_1(Q)^{\flat}(U^x_se_j)\transport^x_{\nabla}(s)\Id W^{x,j}_s-  \int^t_0\transport^x_{\nabla}(s)^{-1}   Q_{\nabla}(\bb_s^x) \transport^x_{\nabla}(s) \Id s,
$$
a.s. on $\{t<\zeta^x\}$, where $e_1,\dots, e_m$ is the standard basis of $\IR^m$, 
$$
U^x:[0,\zeta^x)\times\Omega\longrightarrow O(\IMM)=\bigcup_{x\in \IMM} O(\IR^m,T_x\IMM)
$$
is a horizontal lift of $\bb^x$ with respect to the Levi-Civita connection on $\IMM$ to the principal fiber bundle of orthonormal frames $O(\IMM)\to \IMM$, and 
$$
W^x:=\int_0^{\bullet}\varpi(*\Id U^x_s) :[0,\zeta^x)\times\Omega\longrightarrow \IR^m
$$
is the $\IR^m$-representation of $\bb^x$ (in particular, $W^x$ is a Euclidean Brownian motion), with 
$$
\varpi\in \Omega^1_{C^{\infty}}( O(\IMM),\IR^m),\quad \varpi_u(A):=u^{-1}(T\pi (A_u)),\quad A_u\in T_uO(\IMM),\quad u\in O(\IMM), 
$$
the solder $1$-form of $\pi:O(\IMM)\to  \IMM$. These constructions do not depend on the initial value $U^x_0\in O(\IR^m,T_x\IMM)$. \vspace{1mm}

It is often useful to know for estimates that the processes of the form $\mathcal{Q}^x_{\nabla}$ factor as follows:

\begin{remark}\label{ui} 
a) Let $\alpha\in\Omega^1_{C^\infty}(\IMM,\mathrm{End}(\IEE))$, $V,W\in\Gamma_{C^\infty}(\IMM,\mathrm{End}(\IEE))$ and let
$$
C:[0,\zeta^x)\times\Omega\longrightarrow\mathrm{End}(\IEE_x)
$$
be the solution to
$$
\Id C(t)=-C(t) \transport^x_{\nabla}(t)^{-1}  \big(V(\bb_t^x)+\alpha(\Id\bb_t^x)+W(\bb_t^x)\Id t\big)\transport^x_{\nabla}(t),\quad C(0)=1.
$$ 
Such a $C$ factors as follows: let
$$
A:[0,\zeta^x)\times\Omega\longrightarrow\mathrm{End}(\IEE_x)
$$
be the solution to
$$
\Id A(t)=-A(t) \transport^x_{\nabla}(t)^{-1}  \big(\alpha(\Id\bb_t^x)+W(\bb_t^x)\Id t\big)\transport^x_{\nabla}(t),\quad A(0)=1.
$$ 
Then $A$ is invertible and 
$$
A^{-1}:[0,\zeta^x)\times\Omega\longrightarrow\mathrm{End}(\IEE_x)
$$
is the solution to
$$
\Id A(t)^{-1}= \transport^x_{\nabla}(t)^{-1}  \big(\alpha(\Id\bb_t^x)+W(\bb_t^x)\Id t\big)\transport^x_{\nabla}(t)A(t)^{-1},\quad A(0)^{-1}=1.
$$ 
Let $B$ be the solution to
$$
\Id B(t)=- B(t)A(t)\transport^x_{\nabla}(t)^{-1} V(\bb_t^x) \transport^x_{\nabla}(t)A(t)^{-1}\Id t,\quad B(0)=1.
$$
Then by the It\^{o} product rule we have
\begin{align*}
\Id (B(t)A(t))&= (\Id B(t))A(t)+B(t)\Id A(t)+ \Id B(t)\Id A(t)\\
&=- B(t)A(t)\transport^x_{\nabla}(t)^{-1} V(\bb_t^x) \transport^x_{\nabla}(t)A(t)^{-1}\Id t A(t)\\
&-B(t)A(t) \transport^x_{\nabla}(t)^{-1}  \big(\alpha(\Id\bb_t^x)+W(\bb_t^x)\Id t\big)\transport^x_{\nabla}(t)\\
&+\quad\text{summands containing $\Id t$ and $\Id\bb^x_t$, or $\Id t$ and $\Id t$,}
\end{align*}
so that by uniqueness $C=AB$.\vspace{1mm}

b) As a particular case of the above situation, 
Let
$$
\mathcal{Q}^x_{1,\nabla}:[0,\zeta^x)\times\Omega\longrightarrow\mathrm{End}(\IEE_x)
$$
be the solution to
$$
\Id \mathcal{Q}^x_{1,\nabla}(t)=-\mathcal{Q}^x_{1,\nabla}(t) \transport^x_{\nabla}(t)^{-1}  \sigma_1(Q)^{\flat}(\Id\bb_t^x)\transport^x_{\nabla}(t),\quad \mathcal{Q}^x_{1,\nabla}(0)=1,
$$ 
and let $\mathcal{Q}^x_{2,\nabla}$ be the solution to
$$
\Id \mathcal{Q}^x_{2,\nabla}(t)=- \mathcal{Q}^x_{2,\nabla}(t)\mathcal{Q}^x_{1,\nabla}(t)\transport^x_{\nabla}(t)^{-1} Q_{\nabla}(\bb_t^x) \transport^x_{\nabla}(t)\mathcal{Q}^x_{1,\nabla}(t)^{-1}\Id t,\quad \mathcal{Q}^x_{2,\nabla}(t)=1.
$$
Then we have 
\begin{align}\label{factor}
\mathcal{Q}^x_{\nabla}(t)=\mathcal{Q}^x_{2,\nabla}(t)\mathcal{Q}^x_{1,\nabla}(t). 
\end{align}
\end{remark}

Any differential operator
$$
Q:\Gamma_{C^{\infty}}(\IMM,\IEE)\longrightarrow \Gamma_{C^{\infty}}(\IMM,\IEE)
$$
induces a densely defined sesqui-linear form
\begin{align}\label{form}
 \Gamma_{C^{\infty}_c}(\IMM,\IEE)\times \Gamma_{C^{\infty}_c}(\IMM,\IEE)\ni (\Psi_1,\Psi_2)\longmapsto h^{\nabla}_Q(\Psi_1,\Psi_2):=\left\langle (\nabla^{\dagger}\nabla/2 +Q)\Psi_1,\Psi_2\right\rangle\in\IC
\end{align}
in $\Gamma_{L^2}(\IMM,\IEE)$. In case this form is sectorial it is automatically closable (stemming from a sectorial operator), and we denote the closed operator in $\Gamma_{L^2}(\IMM,\IEE)$ induced by the closure of $h^{\nabla}_Q$ with $H^{\nabla}_Q$ in the sense of Theorem \ref{gene} from the appendix. It follows that $H^{\nabla}_Q$ generates a holomorphic semigroup (cf. appendix)
$$
(\mathrm{e}^{-z H^{\nabla}_Q})_{z\in \Sigma_{0,\beta} }\subset \ILL(\Gamma_{L^2}(\IMM,\IEE)),
$$
which is defined on some sector of the form 
$$
 \Sigma_{0,\beta}=\{r\mathrm{e}^{\sqrt{-1}\alpha}:r\geq 0, \alpha\in (-\beta,\beta)\}\quad\text{ for some $\beta\in (0,\pi/2]$.}
$$

In the situation of a trivial complex line bundle with its trivial connection (identifying sections with functions) we will ommit the dependence on the connection in the notation. In particular, $H\geq 0$ stands for the Friedrichs realization of the scalar Laplace-Beltrami operator $\Delta/2$ in $L^2(\IMM)$.

\begin{theorem}\label{main} Let 
$$
Q:\Gamma_{C^{\infty}}(\IMM,\IEE)\longrightarrow \Gamma_{C^{\infty}}(\IMM,\IEE)
$$
be a differential operator of order $\leq 1$. Assume that $h^{\nabla}_{Q}$ is sectorial and that 
\begin{align}\label{ass2}
\sup_{x\in K} \mathbb{E}\left[1_{\{t<\zeta^x\}}|\mathcal{Q}_{\nabla}^x(t)|^2\right]<\infty\quad\text{for all $K\subset\IMM$ compact, $t>0$.}
\end{align}
%is sectorial 
%$$
%\sigma_1(Q)\in \Gamma_{L^{\infty}}\big(\IMM,\Hom( T^{\dagger}\IMM, \mathrm{End}(\IEE))\big)
%$$
%and that one has the Kato property
%\begin{align}\label{ass3}
%\lim_{t\to 0+}\sup_{x\in\IMM}\int^T_0\mathbb{E}\left[1_{\{t<\zeta^x\}}|Q_{\nabla}(\bb^x_s)| \right]\Id s<1.
%\end{align}
Then for all $t>0$, $\Psi\in \Gamma_{L^2}(\IMM,\IEE)$, $x\in \IMM$, one has 
\begin{align}\label{ass}
\mathrm{e}^{-t H^{\nabla}_Q}\Psi(x) =\mathbb{E}\left[1_{\{t<\zeta^x\}}\mathcal{Q}_{\nabla}^x(t) \transport_{\nabla}^{x}(t)^{-1}\Psi(\mathsf{b}^x_t)\right].
\end{align}
\end{theorem}

\begin{remark} By local parabolic regularity, the time dependent section $(t,x)\mapsto \mathrm{e}^{-t H^{\nabla}_Q}\Psi(x)$ has a representative which is smooth on $(0,\infty)\times \IMM$, and (\ref{ass}) means that the RHS of this equation is precisely this smooth representative. This pointwise identification, which is based on the locally uniform integrability assumption (\ref{ass2}), is highly nontrivial in the stochastically incomplete case and even slightly improves the existing results in the 'usual' Feynman-Kac setting ($\sigma_1(Q)=0$ and $Q_{\nabla}$ self-adjoint), where so far only an $\mu$-almost everywhere equality has been established.  
\end{remark}

\begin{proof}[Proof of Theorem \ref{main}] We omit the dependence on $\nabla$ of several data in the notation, whenever there is no danger of confusion. Fix $x\in\IMM$, $t>0$ and pick an exhaustion $(U_l)_{l\in\IN}$ of $\IMM$ with open connected relatively compact subsets having a smooth boundary. Let $H_{Q,l}$ be defined with $\IMM$ replaced by $U_l$ (note that this corresponds to Dirichlet boundary conditions). It suffices to show that (with an obvious notation) for all $\Psi\in \Gamma_{C^{\infty}_c}(\IMM,\IEE)$ and all $l$ large enough such that $\Psi$ is supported in $U_l$ one has
\begin{align}\label{reicht}
\mathrm{e}^{-t H_{Q,l}}\Psi(x) =\mathbb{E}\left[1_{\{t<\zeta^l_x\}}\mathcal{Q}^x(t) \transport^{x}(t)^{-1}\Psi(\mathsf{b}^x_t)\right].
\end{align}
Indeed, we have
\begin{align}\label{ernst}
\lim_{l\to\infty}\left\|\mathrm{e}^{-tH_{Q,l}}\Psi- \mathrm{e}^{-tH_Q}\Psi\right\|_2=0
\end{align}
by an abstract monotone convergence theorem for nondensely defined sectorial forms (Theorem 3.7 in \cite{chill}), and furthermore for every compact set $K\subset \IMM$ with $x\in K$ we have
\begin{align*}
&\sup_{y\in K}\left|\mathbb{E}\left[(1_{\{t<\zeta^y\}}-1_{\{t<\zeta_l^y\}})\mathcal{Q}^y(t) \transport^{y}(t)^{-1}\Psi(\mathsf{b}^y_t)\right]\right|\\
&\leq \sup_{y\in K} \left\|\Psi\right\|_{\infty}\mathbb{E}\left[1_{\{t<\zeta^y\}}-1_{\{t<\zeta_l^y\}}\right]^{1/2}\mathbb{E}\left[1_{\{t<\zeta^y\}}|\mathcal{Q}^{y}(t)|^2 \right]^{1/2}\\
&\leq \sup_{y\in K}\mathbb{E}\left[1_{\{t<\zeta^y\}}|\mathcal{Q}^{y}(t)|^2 \right]^{1/2} 2^{1/2} \sup_{y\in K}\left\|\Psi\right\|_{\infty}(\mathrm{e}^{-tH}1(y)-\mathrm{e}^{-tH_l}1(y))^{1/2}.
\end{align*}
The latter expression converges to zero as $l\to\infty$ by a maximum principle for the heat equation of Dodziuk \cite{dod}, which shows that the RHS of (\ref{ass}) is continuous in $x$, and that in view of (\ref{ernst}) one has (\ref{ass}) for $\mu$-a.e $x\in \IMM$. A posteriori this equality holds for \emph{all $x$}, as both sides are continuous in $x$. If $\Psi$ is only square integrable, we can pick a sequence of smooth compactly supported sections $(\Psi_n)_{n\in\IN}$ with $\left\|\Psi_n-\Psi\right\|_2\to 0$. Given an open relatively compact subset $U\subset \IMM$ with $x\in U$, we have
$$
\mathrm{e}^{-tH_{Q}}:\Gamma_{L^2}(\IMM,\IEE)\longrightarrow \Gamma_{C_b}(U,\IEE)
$$
\emph{algebraically} by elliptic regularity (where $\Gamma_{C_b}(U,\IEE)$ denotes the Banach space of continuous bounded sections of $\IEE|_{U}\to U$ equipped with the uniform norm), and a posteriori \emph{continuously} by the closed graph theorem, we then have 
$$
\lim_{n\to\infty}\mathrm{e}^{-tH_Q}\Psi_n(x)=\mathrm{e}^{-tH_Q}\Psi(x),
$$
and
\begin{align*}
&\left|\mathbb{E}\left[1_{\{t<\zeta^x\}}\mathcal{Q}^x(t) \transport^{x}(t)^{-1}(\Psi_n(\mathsf{b}^x_t)-\Psi(\mathsf{b}^x_t))\right]\right|\\
&\leq \mathbb{E}\left[1_{\{t<\zeta^x\}}|\mathcal{Q}^x(t)|^2\right]^{1/2}\mathbb{E}\left[1_{\{t<\zeta^x\}}|\Psi_n(\mathsf{b}^x_t)-\Psi(\mathsf{b}^x_t)|^2\right]^{1/2}\\
&= \mathbb{E}\left[1_{\{t<\zeta^x\}}|\mathcal{Q}^x(t)|^2\right]^{1/2}\left(\int \mathrm{e}^{-tH}(x,y)|\Psi_n(y)-\Psi(y)|^2\Id\mu(y)\right)^{1/2}\\
&\leq \mathbb{E}\left[1_{\{t<\zeta^x\}}|\mathcal{Q}^x(t)|^2\right]^{1/2}\left(\sup_{y\in\IMM}\mathrm{e}^{-tH}(x,y)\right)^{1/2}\left\|\Psi_n-\Psi\right\|_2,
\end{align*}
which tends to $0$ as $n\to\infty$ and proves (\ref{ass}) again.\\
It remains to show (\ref{reicht}): By parabolic regularity, the time dependent section 
$$
\Psi_s(y):=\mathrm{e}^{-(t-s)H_{Q,l}}\Psi(y) 
$$
of $\IEE|_{U_l}\to U_l$ extends smoothly to $[0,t]\times \overline{U_l}$ and $\Psi_s$ vanishes in $\partial U_l$ for all $s\in [0,t)$. Define a continuous semimartingale by
$$
N:[0,t\wedge \zeta^x_l]\times \Omega\longrightarrow \IEE_x,\quad N_s:=\mathcal{Q}^x(s) \transport^{x}(s)^{-1}\Psi_s(\bb_s^x).
$$
Then we have 
\begin{align*}
&\Id  N_s= (\Id\mathcal{Q}^x(s))\transport^{x}(s)^{-1}\Psi_s(\bb_s^x)+\mathcal{Q}^x(s)\Id\transport^{x}(s)^{-1}\Psi_s(\bb_s^x)+\Id\mathcal{Q}^x(s)\Id\transport^{x}(s)^{-1}\Psi_s(\bb_s^x)\\
&=-\mathcal{Q}^x(s)\transport^x (s)^{-1} \big( \sigma_1(Q)^{\flat}(\Id\bb_s^x)+ Q_{\nabla}(\bb_s^x) \Id s\big)\Psi_s(\bb_s^x)\\
&\quad+\mathcal{Q}^x(s)\left(\transport^{x}(s)^{-1}\nabla\Psi_s(*\Id\bb_s^x)(\bb_s^x)+\transport^{x}(s)^{-1}\partial_s\Psi_s(\bb_s^x)\Id  s\right)\\
&\quad-\mathcal{Q}^x(s)\transport^x (s)^{-1} \big( \sigma_1(Q)^{\flat}(\Id\bb_s^x)+ Q_{\nabla}(\bb_s^x) \Id s\big)\transport^x (s)\\
&\quad\quad\times\left(\transport^{x}(s)^{-1}\nabla\Psi_s(*\Id\bb_s^x)(\bb_s^x)+\transport^{x}(s)^{-1}\partial_s\Psi_s(\bb_s^x)\Id s\right)\\
&\equiv -\mathcal{Q}^x(s)\transport^x (s)^{-1} \big(  Q_{\nabla}(\bb_s^x) \Id s\big)\Psi_s(\bb_s^x)\\
&\quad+\mathcal{Q}^x(s)\left(\transport^{x}(s)^{-1}\nabla\Psi_s(*\Id\bb_s^x)(\bb_s^x)-\frac{1}{2}\transport^{x}(s)^{-1}\nabla^{\dagger}\nabla\Psi_s(\bb_s^x)\Id s+\transport^{x}(s)^{-1}\partial_s\Psi_s(\bb_s^x)\Id s\right)\\
&\quad-\mathcal{Q}^x(s)\transport^x (s)^{-1} \big( \sigma_1(Q)^{\flat}(\Id\bb_s^x)+ Q_{\nabla}(\bb_s^x) \Id s\big)\transport^x (s)\\
&\quad\quad\times\left(\transport^{x}(s)^{-1}\nabla\Psi_s(*\Id\bb_s^x)(\bb_s^x)+\frac{1}{2}\transport^{x}(s)^{-1}\nabla^{\dagger}\nabla\Psi_s(\bb_s^x)\Id s+\transport^{x}(s)^{-1}\partial_s\Psi_s(\bb_s^x)\Id s\right)\\
&\equiv -\mathcal{Q}^x(s)\transport^x (s)^{-1}  Q_{\nabla}(\bb_s^x) \Id s\Psi_s(\bb_s^x)+\mathcal{Q}^x(s)\left(\frac{-1}{2}\transport^{x}(s)^{-1}\nabla^{\dagger}\nabla\Psi_s(\bb_s^x)\Id s+\transport^{x}(s)^{-1}\partial_s\Psi_s(\bb_s^x)\Id  s\right)\\
&\quad-\mathcal{Q}^x(s)\transport^x (s)^{-1} \big( \sigma_1(Q)^{\flat}(\Id\bb_s^x)+ Q_{\nabla}(\bb_s^x) \Id s\big)\transport^x (s)\\
&\quad\quad\times\left(\transport^{x}(s)^{-1}\nabla\Psi_s(*\Id\bb_s^x)(\bb_s^x)+\frac{-1}{2}\transport^{x}(s)^{-1}\nabla^{\dagger}\nabla\Psi_s(\bb_s^x)\Id s+\transport^{x}(s)^{-1}\partial_s\Psi_s(\bb_s^x)\Id s\right)\\
&= -\mathcal{Q}^x(s)\transport^x (s)^{-1}   Q_{\nabla}(\bb_s^x) \Id s\Psi_s(\bb_s^x)+\mathcal{Q}^x(s)\left(\frac{-1}{2}\transport^{x}(s)^{-1}\nabla^{\dagger}\nabla\Psi_s(\bb_s^x)+\transport^{x}(s)^{-1}\partial_s\Psi_s(\bb_s^x)\Id s\right)\\
&\quad-\mathcal{Q}^x(s)\transport^x (s)^{-1} \sigma_1(Q)^{\flat}(\Id\bb_s^x)\nabla\Psi_s(*\Id\bb_s^x)(\bb_s^x)\\
&=- \mathcal{Q}^x(s)\transport^x (s)^{-1}   Q_{\nabla}(\bb_s^x) \Id s\Psi_s(\bb_s^x)+\mathcal{Q}^x(s)\left(\frac{-1}{2}\transport^{x}(s)^{-1}\nabla^{\dagger}\nabla\Psi_s(\bb_s^x)+\transport^{x}(s)^{-1}\partial_s\Psi_s(\bb_s^x)\Id s\right)\\
&\quad-\mathcal{Q}^x(s)\transport^x (s)^{-1}  \sigma_1(Q)\nabla\Psi_s(\bb_s^x)\\
&=0,
\end{align*}
where $\equiv$ stands for equality up to continuous local martingales. In the above calculation, we have used the It\^{o} product rule, the differential equation for $\mathcal{Q}^x$, the formula
$$
\Id\transport^{x}(s)^{-1}\Psi_s(\bb_s^x)=\transport^x(s)^{-1}\nabla\Psi_s(*\Id\bb_s^x)(\bb_s^x)+\transport^x(s)^{-1}_s\partial_s\Psi_s(\bb_s^x),
$$
which follows from applying (\ref{para}) to the metric connection $\pi^{*}\nabla$ on the metric vector bundle $\pi^{*}\IEE\to \IMM\times [0,\infty)$ with the projection $\pi:\IMM\times [0,\infty)\to \IMM$, the covariant Stratonovic-to-It\^{o} formula
$$
\transport^{x}(s)^{-1}\nabla\Psi_s(*\Id\bb_s^x)(\bb_s^x)=\transport^{x}(s)^{-1}\nabla\Psi_s(*\Id\bb_s^x)(\bb_s^x)+\frac{1}{2}\transport^{x}(s)^{-1}\nabla^{\dagger}\nabla \Psi_s(\bb_s^x)\Id s,
$$
and 
$$
\partial_s\Psi_s= ((1/2)\nabla^{\dagger}\nabla+\sigma_1(Q)\nabla+Q_{\nabla})\Psi_s.
$$
This shows that $N$ is a continuous local martingale. Since $U_l$ is relatively compact, $N$ is in fact a martingale: indeed, a.s., for all $s>0$ we have in $\{ s<\zeta^x\}$ from the differential equation for $\mathcal{Q}^{x}$ and Jenßen's inequality 
\begin{align*}
&\left|\mathcal{Q}^{x}(s)\right|^2\leq   C+C\left|\int^s_0\mathcal{Q}^x(r)\transport^x(r)^{-1} \sigma_1(Q)^{\flat}(\Id\bb_r^x)\transport^x(r)\right|^2+Cs\int^s_0|\mathcal{Q}^x(r)|^2| Q(\bb_r^x)|^2\Id r,
\end{align*}
so that by the Burkholder-Davis-Gundy inequality, with 
$$
\vartheta_n:=\inf\{r\geq 0:\left|\mathcal{Q}^{x}(r)\right|>n\},\quad n\in\IN,
$$
one has
\begin{align*}
&\mathbb{E}\Big[\sup_{s\leq t\wedge \zeta^x_l}\left|\mathcal{Q}^{x}(s\wedge\vartheta_n)\right|^2\Big]\\
&\leq C'+C'\mathbb{E}\left[    \int^{t\wedge \zeta^x_l}_0|\mathcal{Q}^x(r\wedge\vartheta_n)|^2  | \sigma_1(Q)^{\flat}(\bb^x_r)|^2\Id r\right]+t\mathbb{E}\left[\int^{t\wedge \zeta^x_l}_0|\mathcal{Q}^x(r\wedge\vartheta_n)|^2| Q_{\nabla}(\bb_r^x)|^2\Id r\right]\\
&\leq C'+C'\left(\sup_{y\in U_l}| \sigma_1(Q)^{\flat}(y)|^2\right)\mathbb{E}\left[    \int^{t\wedge \zeta^x_l}_0|\mathcal{Q}^x(r\wedge\vartheta_n)|^2  \Id r\right]\\
&\quad\quad+t\left(\sup_{y\in U_l}| Q_{\nabla}(y)|^2\right)\mathbb{E}\left[\int^{t\wedge \zeta^x_l}_0|\mathcal{Q}^x(r\wedge\vartheta_n)|^2\Id r\right]\\
&\leq C_{Q,l}+(C_{Q,l}+tC_{Q,l})\mathbb{E}\left[\int^{t\wedge \zeta^x_l}_0|\mathcal{Q}^x(r\wedge\vartheta_n)|^2\Id r\right]\\
&\leq C_{Q,l}+(C_{Q,l}+tC_{Q,l})\mathbb{E}\left[\int^{t}_0\sup_{s\leq r\wedge \zeta^x_l}|\mathcal{Q}^x(s\wedge\vartheta_n)|^2\Id r\right],
\end{align*}
where $C$, $C'$ are universal constants, and $C_{Q,l} $ depends only on $\left\|Q_{\nabla}|_{U_l}\right\|_{\infty}$ and $\left\|\sigma_1(Q)|_{U_l}\right\|_{\infty}$. As a consequence, for all $T>0$ with $t\leq T$, Gronwall's inequality gives
$$
\mathbb{E}\left[\sup_{s\leq t\wedge \zeta^x_l}\left|\mathcal{Q}^{x}(s\wedge\vartheta_n)\right|^2\right]\leq C_{Q,l} \mathrm{e}^{C_{Q,l,T}t}, 
$$
where $C_{Q,l,T}$ only depends on $Q,l,T$, and so
\begin{align}\label{absch}
&\mathbb{E}\left[\sup_{s\leq t\wedge \zeta^x_l}\left|\mathcal{Q}^{x}(s)\right|^2\right]=\mathbb{E}\left[\max_{s\leq t\wedge \zeta^x_l}\left|\mathcal{Q}^{x}(s)\right|^2\right]=\mathbb{E}\left[\lim_n\max_{s\leq t\wedge \zeta^x_l}\left|\mathcal{Q}^{x}(s\wedge\vartheta_n)\right|^2\right]\\
&\leq \liminf_n\mathbb{E}\left[\sup_{s\leq t\wedge \zeta^x_l}\left|\mathcal{Q}^{x}(s\wedge\vartheta_n)\right|^2\right]\leq C_{Q,l} \mathrm{e}^{C_{Q,l,T}t}<\infty 
\end{align}
by Fatou's lemma. We arrive at 
\begin{align*}
\mathbb{E}\left[\sup_{s\leq t\wedge \zeta^x_l}\left|N_s\right|^2\right]\leq \left(\sup_{s\in [0,t],y\in U_l}|\Psi_s(y)|^2\right)  \mathbb{E}\left[\sup_{s\leq t\wedge \zeta^x_l}\left|\mathcal{Q}^{x}(s)\right|^2\right]<\infty,
\end{align*}
so that
$$
\mathbb{E}\left[\sup_{s\leq t\wedge \zeta^x_l}\left|N_s\right|\right]\leq \mathbb{E}\left[\sup_{s\leq t\wedge \zeta^x_l}\left|N_s\right|^2\right]^{1/2}<\infty,
$$
which shows that $N$ is a martingale, as claimed.\\
We thus have
\begin{align*}
&\mathrm{e}^{-tH^l_Q}\Psi(x)=\mathbb{E}[N_0]=\mathbb{E}[N_{t\wedge \zeta^x_l}]\\
&=\mathbb{E}\left[\mathcal{Q}^x(t\wedge \zeta^x_l) \transport^{x}(t\wedge \zeta^x_l)^{-1}\Psi_{t\wedge \zeta^x_l}(\bb_{t\wedge \zeta^x_l}^x)\right]\\
&=\mathbb{E}\left[(1_{\{t<\zeta^l_x\}}+1_{\{t\geq \zeta^l_x\}})\mathcal{Q}^x(t\wedge \zeta^x_l) \transport^{x}(t\wedge \zeta^x_l)^{-1}\Psi_{t\wedge \zeta^x_l}(\bb_{t\wedge \zeta^x_l}^x)\right]\\
&=\mathbb{E}\left[1_{\{t<\zeta^l_x\}}\mathcal{Q}^x(t) \transport^{x}(t)^{-1}\Psi_{t\wedge \zeta^x_l}(\bb_{t}^x)\right]+\mathbb{E}\left[1_{\{t\geq \zeta^l_x\}}\mathcal{Q}^x( \zeta^x_l) \transport^{x}(\zeta^x_l)^{-1}\Psi_{t\wedge \zeta^x_l}(\bb_{ \zeta^x_l}^x)\right]\\
&=\mathbb{E}\left[1_{\{t<\zeta^l_x\}}\mathcal{Q}^x(t) \transport^{x}(t)^{-1}\Psi(\mathsf{b}^x_t)\right].
\end{align*}
This completes the proof.
\end{proof}

In order to evaluate the somewhat abstract assumptions from Theorem \ref{main}, we recall the definition of the Kato class (referring the reader to \cite{batu2, peter, aizen, semigroup, batuvil,sturm2} and the refernces therein for some fundamental results concerning this class):

\begin{definition}\label{QQQ} A Borel function $w:\IMM\to \IR$ is said to be in the Kato class $\mathcal{K}(\IMM)$ of $\IMM$, if
$$
\lim_{t\to 0+}\sup_{x\in\IMM}\int^t_0 \mathbb{E}\left[1_{\{s<\zeta^x\}}|w(\bb_s^x)|\right] \Id s=0.
$$
\end{definition}

By Khashminskii's lemma \cite{batu2}, $w\in \mathcal{K}(\IMM)$ implies
$$
\sup_{x\in\IMM}\mathbb{E}\left[1_{\{t<\zeta^x\}}\mathrm{e}^{p\int^t_0|w(\bb_s^x)|\Id s}\right]<\infty\quad\text{for all $t>0$, $p\in [1,\infty)$.}
$$
One trivially always has $L^{\infty}(\IMM)\subset \mathcal{K}(\IMM)$, and under a mild control on the geometry one has $L^p+L^{\infty}$-type subspaces of the Kato class. For example, one has (cf. Chapter VI in \cite{batu2} and the appendix of \cite{braun}):

\begin{proposition}\label{PPP}a) Assume there exists a Borel function $\theta:\IMM\to (0,\infty)$ with
$$
\sup_{x\in\IMM} \mathrm{e}^{-t H}(x,y)\leq  \theta(y)t^{-m/2}\quad\text{ for all $0<t<1$, $y\in\IMM$.}
$$
Then one has 
$$
L^{p}_{\theta}(\IMM)+L^{\infty}(\IMM)\subset \mathcal{K}(\IMM),\quad \text{for all $p\geq 1$ if $m=1$, and all $p>m/2$ if $m\geq 2$,}
$$
where $L^{p}_{\theta}(\IMM)$ denotes the weighted $L^p$-space of all equivalence classes of Borel functions $f$ on $\IMM$ such that $\int |f|^p \theta \Id\mu<\infty$. \\
b) If $\IMM$ is geodesically complete and quasi-isometric to a Riemannian manifold with Ricci curvature bounded from below by a constant, then one has
$$
L^{p}_{1/\mu(B(\cdot,1))}(\IMM)+L^{\infty}(\IMM)\subset \mathcal{K}(\IMM),\quad \text{for all $p\geq 1$ if $m=1$, and all $p>m/2$ if $m\geq 2$.}
$$
\end{proposition}

Given an endomorphism $A$ on a metric vector bundle, we denote with 
$$
\Re(A):=(1/2)(A+A^{\dagger})
$$
its real part and with 
$$
\Im(A):=-\sqrt{-1}(A-\Re(A))
$$
its imaginary part, so that $A=\Re(A)+\sqrt{-1}\Im(A)$, where $\Re(A)$ and $\Im(A)$ are self-adjoint (and then, for example, the positive and negative parts $\Re(A)^{\pm}\geq 0$ are defined via the fiberwise spectral calculus, giving $\Re(A)=\Re(A)^{+}-\Re(A)^{-}$). Note also that $\Re(A)=\Re(A^{\dagger})$, and that $\Re(A)=U\Re(B)U^{\dagger}$ if $A=UBU^{\dagger}$ for some unitary $U$.

\begin{proposition}\label{aaspopo} Let 
$$
Q:\Gamma_{C^{\infty}}(\IMM,\IEE)\longrightarrow \Gamma_{C^{\infty}}(\IMM,\IEE)
$$
be a differential operator of order $\leq 1$.\\
a) Assume 
\begin{itemize} 
\item $|\Re(\sigma_1(Q))|\in L^{\infty}(\IMM)$, 
\item $\Re(Q_{\nabla})$ is bounded from below by a constant $\kappa\in\IR$,
\item $|\Im(Q_{\nabla})|\in\mathcal{K}(\IMM)$.
\end{itemize}
Then $h^{\nabla}_{Q}$ is sectorial and 
\begin{align}\label{ass3}
\sup_{x\in \IMM} \mathbb{E}\left[1_{\{t<\zeta^x\}}|\mathcal{Q}_{\nabla}^x(t)|^2\right]<\infty\quad\text{for all $t>0$},
\end{align}
in particular, (\ref{ass}) holds true.\\
b) Assume 
\begin{itemize} 
\item $\sigma_1(Q)$ is anti-selfadjoint and $|\sigma_1(Q)|\in L^{\infty}(\IMM)$,
\item $|\Re(Q_{\nabla})^-|\in \mathcal{K}(\IMM)$,
\item $|\Im(Q_{\nabla})|\in \mathcal{K}(\IMM)$.
\end{itemize}
Then $h^{\nabla}_{Q}$ is sectorial and one has (\ref{ass3}), in particular, (\ref{ass}) holds true.
\end{proposition}

\begin{proof} We have 
$$
h^{\nabla}_{Q}= h_a+h_b+h_c+h_d+h_e,
$$
where 
\begin{align*}
&h_a(\Psi_1,\Psi_2):=(1/2) \left\langle \nabla \Psi_1,\nabla\Psi_2\right\rangle,\quad h_b(\Psi_1,\Psi_2):= \left\langle  \sigma_1(Q)\nabla\Psi_1,\Psi_2\right\rangle,\\
& h_c(\Psi_1,\Psi_2):= \left\langle \Re(Q_{\nabla})^+\Psi_1,\Psi_2\right\rangle,\quad h_d(\Psi_1,\Psi_2):=\left\langle \Re(Q_{\nabla})^-\Psi_1,\Psi_2\right\rangle,\\
& h_e(\Psi_1,\Psi_2):=\left\langle \Im(Q_{\nabla})\Psi_1,\Psi_2\right\rangle.
\end{align*} 
a) We have
\begin{align}
|h_b(\Psi,\Psi)|\leq \left\|\sigma_1(Q)\right\|_{\infty}  \left\|\nabla\Psi\right\|\left\|\Psi\right\|\leq \left\|\sigma_1(Q)\right\|_{\infty}  \left(C_\epsilon\left\|\Psi\right\|^2+ \epsilon h_a(\Psi,\Psi)\right),
\end{align}
and (as Kato perturbations of Bochner-Laplacians are infinitesimally form small; cf. Lemma VII.4 in \cite{batu2})
$$
|h_e(\Psi,\Psi)|\leq \left(C_\epsilon\left\|\Psi\right\|^2+  \epsilon h_a(\Psi,\Psi)\right),
$$
which shows that $h_a+h_b+h_e$ is sectorial, as $h_a$ is so (cf. Theorem \ref{klmn} in the appendix). Moreover,
$$
h_c(\Psi,\Psi)+h_d(\Psi,\Psi)= \left\langle \Re(Q_{\nabla})\Psi,\Psi\right\rangle
$$
is bounded from below, so that the sum 
$$
h=h_a+h_b+h_e+h_c+h_d
$$
of sectorial forms is sectorial, too. \\
Let $v\in\IEE_x$. Almost surely, for all $s>0$, we have in $\{s<\zeta^x\}$ by the It\^{o} product rule,
\begin{align*}
&\Id \left| \mathcal{Q}_{\nabla}^x(s )^{\dagger}v\right|^2= 2\Re\left(\Id \mathcal{Q}_{\nabla}^x(s)^{\dagger}v,\mathcal{Q}_{\nabla}^x(s)^{\dagger}v\right)+\left(\Id \mathcal{Q}_{\nabla}^x(s)^{\dagger}v,\Id \mathcal{Q}_{\nabla}^x(s)^{\dagger}v\right)\\
&\leq -2\big( \transport^x_{\nabla}(s)^{-1}  \Re(\sigma_1(Q)^{\flat}(\Id\bb_{s}^x))\transport^x_{\nabla}(t)\mathcal{Q}_{\nabla}^x(s)^{\dagger}v,\mathcal{Q}_{\nabla}^x(s)^{\dagger}v\big)\\
&\quad -2\left( \transport^x_{\nabla}(s)^{-1} \Re(Q_{\nabla}(\bb_{s}^x)) \transport^x_{\nabla}(s)\mathcal{Q}_{\nabla}^x(s)^{\dagger}v,\mathcal{Q}_{\nabla}^x(s)^{\dagger}v\right) \Id s\\
&\quad+|\sigma_1(Q)^{\flat}(\bb_{s}^x)|^2|Q_{\nabla}(s)^{\dagger}v|^2\Id s\\
&\leq -2\left( \transport^x_{\nabla}(s)^{-1}  \Re(\sigma_1(Q)^{\flat}(\Id\bb_{s}^x))\transport^x_{\nabla}(s)\mathcal{Q}_{\nabla}^x(s)v,Q_{\nabla}(s)^{\dagger}v\right)\\
&\quad -2 \kappa |\mathcal{Q}_{\nabla}^x(s)^{\dagger}v|^2 \Id s\\
&\quad+\left\|\Re(\sigma_1(Q))\right\|^2_{\infty}|\mathcal{Q}_{\nabla}^x(s)^{\dagger}v|^2\Id s. 
 \end{align*}
With the sequences of stopping times $\vartheta_n$ and $\zeta^x_l$ as in the proof of Theorem \ref{main}, the It\^{o} isometry an Jenßen's inequality imply that for all $t>0$,
\begin{align*}
&\mathbb{E}\left[\left| \mathcal{Q}_{\nabla}^x(t\wedge\vartheta_n\wedge\zeta^x_l)^{\dagger}v\right|^2\right]\\
&\leq 1+ 2\mathbb{E}\left[\left|\int^t_0\left( \transport^x_{\nabla}(r)^{-1}  \Re(\sigma_1(Q)^{\flat}(\Id\bb_{r}^x))\transport^x_{\nabla}(r)\mathcal{Q}_{\nabla}^x(r)^{\dagger}v,Q_{\nabla}^x(r)^{\dagger}v\right)\right|^{2\frac{1}{2}}|_{r=s\wedge\vartheta_n\wedge\zeta^x_l}\right]\\
&\quad\quad-2\kappa \int^t_0\mathbb{E}\left[ |\mathcal{Q}_{\nabla}^x(s\wedge\vartheta_n\wedge\zeta^x_l)^{\dagger}v|^2 \right]\Id s +\left\|\Re(\sigma_1(Q))\right\|^2_{\infty}\int^t_0\mathbb{E}\left[|\mathcal{Q}_{\nabla}^x(s\wedge\vartheta_n\wedge\zeta^x_l)^{\dagger}v|^2\right] \Id s\\
&\leq 1+ 2 \left\|\Re(\sigma_1(Q))\right\|_{\infty}\mathbb{E}\left[\int^t_0|\mathcal{Q}_{\nabla}^x(s\wedge\vartheta_n\wedge\zeta^x_l)^{\dagger}v|^2\Id s\right]^{\frac{1}{2}}\\
&\quad\quad-2\kappa \int^t_0\mathbb{E}\left[|\mathcal{Q}_{\nabla}^x(s\wedge\vartheta_n\wedge\zeta^x_l)^{\dagger}v|^2 \right]\Id s +\left\|\Re(\sigma_1(Q))\right\|^2_{\infty}\int^t_0\mathbb{E}\left[|\mathcal{Q}_{\nabla}^x(s\wedge\vartheta_n\wedge\zeta^x_l)^{\dagger}v|^2\right] \Id s\\
&\leq 1+ 2 \left\|\Re(\sigma_1(Q))\right\|_{\infty}\left(\mathbb{E}\left[\int^t_0|\mathcal{Q}_{\nabla}^x(s\wedge\vartheta_n\wedge\zeta^x_l)^{\dagger}v|^2\Id s\right]+1\right)\\
&\quad\quad-2\kappa \int^t_0\mathbb{E}\left[ |\mathcal{Q}_{\nabla}^x(s\wedge\vartheta_n\wedge\zeta^x_l)^{\dagger}v|^2 \right]\Id s +\left\|\Re(\sigma_1(Q))\right\|^2_{\infty}\int^t_0\mathbb{E}\left[|\mathcal{Q}_{\nabla}^x(s\wedge\vartheta_n\wedge\zeta^x_l)^{\dagger}v|^2\right] \Id s\\
&\leq 1+ 2 \left\|\Re(\sigma_1(Q))\right\|_{\infty}+2 \left\|\Re(\sigma_1(Q))\right\|_{\infty}\mathbb{E}\left[\int^t_0|\mathcal{Q}_{\nabla}^x(s\wedge\vartheta_n\wedge\zeta^x_l)^{\dagger}v|^2\Id s\right]\\
&-2\kappa \int^t_0\mathbb{E}\left[ |\mathcal{Q}_{\nabla}^x(s\wedge\vartheta_n\wedge\zeta^x_l)^{\dagger}v|^2 \right]\Id s +\left\|\Re(\sigma_1(Q))\right\|^2_{\infty}\int^t_0\mathbb{E}\left[|\mathcal{Q}_{\nabla}^x(s\wedge\vartheta_n\wedge\zeta^x_l)^{\dagger}v|^2\right] \Id s.
\end{align*}
By Gronwall's lemma and Fatou's lemma, this estimate implies
\begin{align*}
&\mathbb{E}\left[1_{\{t<\zeta^x\}}\left| \mathcal{Q}_{\nabla}^x(t)^{\dagger}v\right|^2\right]\leq\lim_l \mathbb{E}\left[1_{\{t<\zeta^x_l\}}\left| \mathcal{Q}_{\nabla}^x(t)^{\dagger}v\right|^2\right]\\
&=\lim_l \mathbb{E}\left[1_{\{t<\zeta^x_l\}}\left| \mathcal{Q}_{\nabla}^x(t\wedge \zeta^x_l)^{\dagger}v\right|^2\right]\leq\lim_l\lim_n \mathbb{E}\left[1_{\{t<\zeta^x_l\}}\left| \mathcal{Q}_{\nabla}^x(t\wedge \vartheta_n\wedge\zeta^x_l)^{\dagger}v\right|^2\right]\\
&\leq C_Q\mathrm{e}^{tC_Q}<\infty,
\end{align*}
uniformly in $x\in\IMM$.\vspace{1mm}

b) As in the proof of part a),
\begin{align*}
|h_b(\Psi,\Psi)|\leq \left\|\sigma_1(Q)\right\|_{\infty}  \left(C_\epsilon\left\|\Psi\right\|^2+ \epsilon h_a(\Psi,\Psi)\right),
\end{align*}
and 
\begin{align*}
&|h_d(\Psi,\Psi)|\leq \left(C_\epsilon\left\|\Psi\right\|^2+\epsilon h_a(\Psi,\Psi)\right),\\
&|h_e(\Psi,\Psi)|\leq \left(C_\epsilon\left\|\Psi\right\|^2+ \epsilon h_a(\Psi,\Psi)\right),
\end{align*}
which shows that $h_a+h_b+h_d+h_e$ is sectorial, and $h_c$ is nonnegative so that $h$ is sectorial.\\
In the notation of Remark \ref{ui}, a.s., for all $s>0$ we have in $\{s<\zeta^x\}$,
$$
\Id \mathcal{Q}^x_{1,\nabla}(s)^{-1}= \transport^x_{\nabla}(s)^{-1}  \sigma_1(Q)^{\flat}(\Id\bb_s^x)\transport^x_{\nabla}(s)\mathcal{Q}^x_{1,\nabla}(s)^{-1},\quad \mathcal{Q}^x_{1,\nabla}(0)^{-1}=1,
$$ 
and
$$
\Id \mathcal{Q}^x_{1,\nabla}(s)^{*}=- \transport^x_{\nabla}(s)^{-1}  \sigma_1(Q)^{\flat}(\Id\bb_s^x)^{\dagger}\transport^x_{\nabla}(s)\mathcal{Q}^x_{1,\nabla}(s)^{*},\quad \mathcal{Q}^x_{1,\nabla}(0)^{*}=1,
$$
which shows that $\mathcal{Q}^x_{1,\nabla}(s)$ is unitary, if $\sigma_1(Q)$ is anti-selfadjoint. Thus we have 
$$
|\mathcal{Q}^x_{\nabla}(s)|=|\mathcal{Q}^x_{2,\nabla}(s)\mathcal{Q}^x_{1,\nabla}(s)|\leq |\mathcal{Q}^x_{2,\nabla}(s)|.
$$
For all $v\in\IEE_x$ (as both $\mathcal{Q}^x_{1,\nabla}(s)$ and the parallel transport are unitary), 
\begin{align*}
&(\Id/\Id s) \left|\mathcal{Q}^x_{2,\nabla}(s)^{\dagger}v\right|^2= 2\Re \Big((\Id/\Id s)\mathcal{Q}^x_{2,\nabla}(s)^{\dagger}v,\mathcal{Q}^x_{2,\nabla}(s)^{\dagger}v\Big) \\
&=-2\Re \Big(   \mathcal{Q}^x_{1,\nabla}(s)\transport^x_{\nabla}(s)^{-1} Q_{\nabla}(\bb_s^x)^{\dagger} \transport^x_{\nabla}(s)\mathcal{Q}^x_{1,\nabla}(s)^{-1}  \mathcal{Q}^x_{2,\nabla}(s)^{\dagger}  v,\mathcal{Q}^x_{2,\nabla}(s)^{\dagger}v\Big)\\
&=-2 \Big(   \mathcal{Q}^x_{1,\nabla}(s)\transport^x_{\nabla}(s)^{-1} \Re(Q_{\nabla}(\bb_s^x)) \transport^x_{\nabla}(s)\mathcal{Q}^x_{1,\nabla}(s)^{-1}  \mathcal{Q}^x_{2,\nabla}(s)^{\dagger}  v,\mathcal{Q}^x_{2,\nabla}(s)^{\dagger}v\Big)\\
&\leq 2 |\Re(Q_{\nabla}(\bb_s^x))^-| |\mathcal{Q}^x_{2,\nabla}(s)^{\dagger}  v|^2
\end{align*}
and so by Gronwall, a.s., for all $t>0$ we have in $\{t<\zeta^x\}$,
$$
|\mathcal{Q}^x_{2,\nabla}(t)|^2=|\mathcal{Q}^x_{2,\nabla}(t)^{\dagger}|^2\leq \mathrm{e}^{2\int^t_0|\Re(Q_{\nabla}(\bb_s^x))^-|\Id s}
$$
and finally
$$
\sup_{x\in\IMM}\mathbb{E}\left[1_{\{t<\zeta^x\}}\mathrm{e}^{2\int^t_0|\Re(Q_{\nabla}(\bb_s^x))^-|\Id s}\right]<\infty
$$
by Khashiminskii's lemma.
\end{proof}

%let  
 %$$
%\bb^{x,y,t}:[0,t]\times\Omega\longrightarrow\IMM
%$$
%be a Brownian bridge $\IMM$ from $x$ to $y$, and let 
%$$
%\transport^{x,y,t}_{\nabla}:[0,t]\times\Omega\longrightarrow  \IEE\boxtimes \IEE^{\dagger}
%$$
%be the corresponding stochastic parallel transport (note here that the Brownian bridge is a continuous semimartingale on $[0,t]$, regardless of the geometry of $\IMM$). Analogously to the Brownian motion case, $\transport^{x,y,t}_{\nabla}$ the uniquely determined continuous semimartingale such that for all $s\in [0,t]$:
%\begin{itemize}
%\item one has $\transport_{\nabla}^{x,y,t}(s):\IEE_x\to\IEE_{\bb_s(x)}$ unitarily, 
%\item for all $\Psi\in \Gamma_{C^{\infty}}(\IMM,\IEE)$ one has
%$$
%\transport_{\nabla}^{x,y,t}(s)^{-1}\Psi(\bb_s^x)= \transport_{\nabla}^{x,y,t}(s)^{-1}\nabla_{*\Id\bb_s(x)}\Psi(\bb_s^x),\quad \transport_{\nabla}^{x,y,t}(0)=1.
%$$
%\end{itemize}
%Finally, let 
%$$
%\mathcal{Q}^{x,y,t}_{\nabla}:[0,t]\times\Omega\longrightarrow\mathrm{End}(\IEE_x)
%$$
%be the unique solution to the Ito equation
%$$
%\Id\mathcal{Q}_{\nabla}^{x,y,t}(s)= \mathcal{Q}_{\nabla}^{x,y,t}(s)\transport_{\nabla}^{x,y,t}(s)^{-1} \big( \sigma_1(Q)^{\flat}(\Id\bb_s^{x,y,t})+ Q_{\nabla}(\bb_s^{x,y,t}) \Id s\big)\transport^{x,y,t}_{\nabla}(s),\quad \mathcal{Q}_{\nabla}^{x,y,t}(0)=1.
%$$
%
%

Given $x\in\IMM$, let $(\mathbb{P}^{x,y}_t)_{t>0,y\in\IMM}$ be the bridge measures associated with $\bb(x)$: for all $t>0$, $y\in\IMM$, the measure $\mathbb{P}^{x,y}_t$ is the uniquely determined probability measure (cf. \cite{plank}, p. 36) on $\{t<\zeta^x\}$ equipped with the sigma-algebra $\IFF^{\bb^x|_{\{t<\zeta^x\}}}_t$ such that
$$
\mathbb{P}^{x,y}_t(A)=\mathbb{E} \left[1_A\f{p(t-s,\bb^x_{s},y)}{ p(t,x,y)} \right]\quad\text{ for all $0<s<t$, $A\in \IFF^{\bb^x|_{\{s<\zeta^x\}}}_s$.}
$$
This provides a pointwise disintegration of Brownian motion, in the sense that for all $t>0$, $x,y\in \IMM$ one has
\begin{align*}
&\mathbb{P}(A)=\int \mathrm{e}^{-tH}(x,y) \mathbb{P}^{x,y}_t(A) \Id\mu(y)\quad\text{for all $A\in \IFF^{\bb^x}_t\cap\{t<\zeta^x\}$},\\
&\mathbb{P}^{x,y}_t(\bb^x_{t}=y)=1.
\end{align*}
We remark that one has to \emph{locally} complete these probability spaces so that $\mathcal{Q}^x_\nabla(t)$ and $\pa^{x}_{\nabla}(t)$ become $\IFF^{\bb^x|_{\{t<\zeta^x\}}}_t$-measurable (cf. p. 250 in \cite{hackenbroch} for a precise treatment of this issue.) \vspace{1mm}

We immediately get the following consequence of Theorem \ref{main}:

\begin{corollary}\label{main12} In the situation of Theorem \ref{main}, for all $t>0$, $x,y\in\IMM$ one has
\begin{align}\label{ass4}
\mathrm{e}^{-t H^{\nabla}_Q}(x,y) = \int_M\mathrm{e}^{-tH}(x,y)\mathbb{E}^{x,y}_t\left[\mathcal{Q}_{\nabla}^{x}(t)\pa^{x}_{\nabla}(t)^{-1}\right].
\end{align}
\end{corollary}

\begin{remark}\label{ali} The precise meaning of this result is as follows: there exists a unique jointly smooth map
$$
(0,\infty)\times\IMM\times\IMM\ni (t,x,y)\longmapsto \mathrm{e}^{-t H^{\nabla}_Q}(x,y)\in \Hom(\IEE_y,\IEE_x)\in \IEE\boxtimes \IEE^{\dagger}
$$
such that for all $t>0$, $x\in\IMM$, $\Psi\in\Gamma_{L^2}(\IMM,\IEE)$ one has
$$
\int |\mathrm{e}^{-t H^{\nabla}_Q}(x,y)|^2\Id\mu(y)<\infty,\quad \mathrm{e}^{-t H^{\nabla}_Q}\Psi(x)=\int \mathrm{e}^{-t H^{\nabla}_Q}(x,y)\Psi(y) \Id\mu(y), 
$$
(this follows from the proof of Theorem II.1 in \cite{batu2}, where the required self-adjointness and semiboundedness of the operator $\tilde{P}$ is only used to get a semigroup which is holomorphic in a sector of the complex plane which contains $(0,\infty)$), and Corollary \ref{main12} states this map is \emph{pointwise} equal to the RHS of (\ref{ass4}).
\end{remark}

In the following result we assume for simplicity that $\IMM$ is compact, in order to not obscure the algebraic machinery behind its proof, and to guarantee the required trace class property:

\begin{theorem}\label{main2} Assume $\IMM$ is compact. Let $V\in\Gamma_{C^{\infty}}(\IMM,\mathrm{End}(\IEE))$ (considered as a differential operator of order $\leq 1$ in $\IEE\to\IMM)$ and let 
$$
P:\Gamma_{C^{\infty}}(\IMM,\IEE)\longrightarrow \Gamma_{C^{\infty}}(\IMM,\IEE)
$$
be a differential operator of order $\leq 1$ and denote its closure in $\Gamma_{L^{2}}(\IMM,\IEE)$, defined a priori on $\Gamma_{C^{\infty}}(\IMM,\IEE)$, with $P$ again. Then for all $t>0$ the operator
\begin{align}\label{ssss}
\int^t_0\mathrm{e}^{-sH^{\nabla}_{V}}P \mathrm{e}^{-(t-s)H^{\nabla}_{V}} \Id s\in \ILL(\Gamma_{L^{2}}(\IMM,\IEE)),
\end{align}
is given for all $x,y\in\IMM$ by 
\begin{align}
%\label{oci}&\big[\int^t_0\mathrm{e}^{-tH^{\nabla}_{P}}w \mathrm{e}^{-tH^{\nabla}_{P}}\Id s\big]\Psi(x)=...\\
\label{oci2}
&\int^t_0\mathrm{e}^{-sH^{\nabla}_{V}}P \mathrm{e}^{-(t-s)H^{\nabla}_{V}}\Id s \ (x,y)\\\nonumber
&=-\mathrm{e}^{-tH}(x,y)\mathbb{E}^{x,y}_t\left[\mathcal{V}^x_{\nabla}(t)\int^t_0 \transport^x_{\nabla}(s)^{-1} \big( \sigma_1(P)^{\flat}(\Id\bb_s^x)+ P_{\nabla}(\bb_s^x) \Id s\big)\transport^x_{\nabla}(s)\pa^{x}_{\nabla}(t)^{-1}\right], 
\end{align}
in particular, for every $\widetilde{V}\in\Gamma_{C^{\infty}}(\IMM,\mathrm{End}(\IEE))$ one has\small{
\begin{align*}
&\mathrm{Tr}\left(\widetilde{V}\int^t_0\mathrm{e}^{-sH^{\nabla}_{V}}P \mathrm{e}^{-(t-s)H^{\nabla}_{V}}\Id s \right)\\\nonumber
&=-\int_{\IMM}\mathrm{e}^{-tH}(x,x)\mathrm{Tr}_{x}\left(\widetilde{V}(x)\mathbb{E}^{x,x}_t\left[\mathcal{V}^x_{\nabla}(t)\int^t_0 \transport^x_{\nabla}(s)^{-1} \big( \sigma_1(P)^{\flat}(\Id\bb_s^x)+ P_{\nabla}(\bb_s^x) \Id s\big)\transport^x_{\nabla}(s)\pa^{x}_{\nabla}(t)^{-1}\right]\right) \Id\mu(x).
\end{align*}
}
\end{theorem}

This result has to be read as follows: by elliptic regularity, for all $t>0$, the function 
$$
[0,t]\ni s\longmapsto \mathrm{e}^{-sH^{\nabla}_{V}}P\mathrm{e}^{-(t-s)H^{\nabla}_{V}}\Psi \in \Gamma_{L^{2}}(\IMM,\IEE)
$$
is well-defined and continuous, so 
$$
\int^t_0\mathrm{e}^{-sH^{\nabla}_{V}}P \mathrm{e}^{-(t-s)H^{\nabla}_{V}}\Psi\Id s
$$
is well-defined in the sense of $\Gamma_{L^{2}}(\IMM,\IEE)$-valued Riemann integrals. Furthermore, 
$$
\Gamma_{L^{2}}(\IMM,\IEE)\ni\Psi\longmapsto\int^t_0\mathrm{e}^{-sH^{\nabla}_{V}}P \mathrm{e}^{-(t-s)H^{\nabla}_{V}}\Psi \Id s\in \Gamma_{L^{2}}(\IMM,\IEE)
$$
is bounded, and our proof shows that $\int^{\bullet}_0\mathrm{e}^{-sH^{\nabla}_{V}}P\mathrm{e}^{-(\bullet-s)H^{\nabla}_{V}}\Id s$ has a jointly smooth integral kernel in the sense of Remark \ref{ali}, and that this smooth representative is pointwise equal to the RHS of (\ref{oci2}). The asserted trace formula then follows from the fact that if an operator $A_1$ in $\Gamma_{L^{2}}(\IMM,\IEE)$ has a smooth integral kernel and $A_2$ is zeroth order, then $A_2A_1$ has the smooth integral kernel $[A_2A_1](x,y)=A_2(x)A_1(x,y)$ and $A_2A_1$ is trace class (as $\IMM$ is compact) with
$$
\mathrm{Tr}\left(A_2A_1\right)= \int_\IMM \mathrm{Tr}_x(A_2(x)A_1(x,x)) \Id\mu(x),
$$
where $\mathrm{Tr}_x$ denotes the finite dimensional trace on $\mathrm{End}(\IEE_x)$.

\begin{proof}[Proof of Theorem \ref{main2}] Denote with $\Lambda(\IR)=\IR\oplus\Lambda^1(\IR)$ the Grassmann algebra over $\IR$, which is generated by $1\in\IR$ and $\theta\in \Lambda^1(\IR)$. In particular, we have $\theta^2=0$. Given a linear space $\mathscr{A}$, the Berezin integral is the linear map
$$
\int_{\Lambda(\IR)}: \mathscr{A}\otimes \Lambda(\IR)\longrightarrow\mathscr{A}, \quad a+b\theta\longmapsto \int_{\Lambda(\IR)}(a+b\theta)\Id \theta:=b,\quad a,b\in\IAA,
$$
which picks the $\theta$-coefficient. Note that if $\mathscr{A}$ is an associative algebra, then so is $\mathscr{A}\otimes \Lambda(\IR)$. With the differential operator 
$$
V+ P^{\theta}:=V+\theta P :\Gamma_{C^{\infty}}(\IMM,\IEE\otimes \Lambda(\IR))=\Gamma_{C^{\infty}}(\IMM,\IEE)\otimes \Lambda(\IR)\longrightarrow \Gamma_{C^{\infty}}(\IMM,\IEE\otimes \Lambda(\IR)),
$$
of order $\leq 1$, the operator $H^{\nabla}_{V+P^{\theta}}$ in 
$$
\Gamma_{L^2}(\IMM,\IEE\otimes \Lambda(\IR))=\Gamma_{L^2}(\IMM,\IEE)\otimes \Lambda(\IR)
$$
is well-defined and in fact equal to the operator sum $H^{\nabla}_{V}+ P^{\theta}$ (as $\IMM$ is compact). The perturbation series 
$$
\mathrm{e}^{-tH^{\nabla}_{V+P^{\theta}}}= 1+\sum^{\infty}_{j=1}\int_{\{0< t_1<\cdots< t_j< t\}} \mathrm{e}^{-t_1H^{\nabla}_{V}} P^{\theta}\mathrm{e}^{-(t_2-t_1)H^{\nabla}_{V}}P^{\theta}\cdots \mathrm{e}^{-(t-t_j)H^{\nabla}_{V}} \Id t_1\cdots \Id t_n
$$
cancels after $j\geq 2$ because of $\theta^2=0$, and we have
\begin{align}\label{forww}
\int_{\Lambda(\IR)}\mathrm{e}^{-tH^{\nabla}_{V+P^{\theta}}}\Id\theta=\int^t_0\mathrm{e}^{-sH^{\nabla}_{V}}P \mathrm{e}^{-(t-s)H^{\nabla}_{V}} \Id s,
\end{align}
in particular, $\int^{\bullet}_0\mathrm{e}^{-sH^{\nabla}_{V}}P \mathrm{e}^{-(\bullet-s)H^{\nabla}_{V}} \Id s$ has a jointly smooth integral kernel in the sense of Remark \ref{ali}. By Corollary \ref{main12} and Remark \ref{ui} we have
$$
\mathrm{e}^{-tH^{\nabla}_{V+P_{\theta}}}(x,y)= \mathrm{e}^{-tH}(x,y)\mathbb{E}^{x,y}_t\left[\mathcal{V}^x_{\nabla}(t)\mathcal{P}^x_{\theta,\nabla}(t)\pa^{x}_{\nabla}(t)^{-1}\right],
$$
where 
$$
\mathcal{P}^x_{\theta,\nabla}:[0,\zeta^x)\times\Omega\longrightarrow\mathrm{End}(\IEE_x\otimes \Lambda(\IR))
$$
denotes the unique solution of
$$
\Id\mathcal{P}^x_{\theta,\nabla}(t)=- \mathcal{P}_{\theta,\nabla}^x(t)\transport^x_{\nabla}(t)^{-1} \big( \sigma_1(P^{\theta})^{\flat}(\Id\bb_t^x)+ P^{\theta}_{\nabla}(\bb_t^x) \Id t\big)\transport^x_{\nabla}(t),\quad \mathcal{P}_{\theta,\nabla}^x(0)=1.
$$
Because of $\theta^2=0$ the time ordered exponential series 
$$ 
\mathcal{P}^x_{\theta,\nabla}(t)=1+\sum_{j=1}^{\infty} \int_{\{0\leq t_1\leq\cdots\leq t_j\leq t\} }\prod_{i=1}^j\theta\Big(- \transport^x_{\nabla}(t_i)^{-1} \big( \sigma_1(P)^{\flat}(\Id\bb_{t_i}^x)+ P_{\nabla}(\bb_{t_i}^x) \Id t_i\big)\transport^x_{\nabla}(t_i)\Big)
$$
has only two summands, giving
\begin{align*}
&\int_{\Lambda(\IR)}\mathrm{e}^{-tH^{\nabla}_{V+P_{\theta}}}(x,y)\Id\theta\\
&=-\mathrm{e}^{-tH}(x,y)\mathbb{E}^{x,y}_t\left[\mathcal{V}^x_{\nabla}(t)\int^t_0 \transport^x_{\nabla}(s)^{-1} \big( \sigma_1(P)^{\flat}(\Id\bb_s^x)+ P_{\nabla}(\bb_s^x) \Id s\big)\transport^x_{\nabla}(s)\pa^{x}_{\nabla}(t)^{-1}\right],
\end{align*}
which in view of (\ref{forww}) is the claimed formula.
\end{proof}

\section{Applications to noncommutative geometry}\label{aaQQ}

In this section we present an application of Theorem \ref{main2} to recent results concerning an algebraic model given in \cite{gl} for Duistermaat-Heckman localization on the space of smooth loops in a compact Riemannian spin manifold. We refer the reader to \cite{LM89} for details on spin geometry (noting that a brief introduction can also be found in \cite{hsu}).\vspace{1mm}

Assume $\IMM$ is a compact Riemannian spin manifold of even dimension, with $\mathscr{S}\to \IMM$ its spin bundle, which is naturally $\mathbb{Z}_2$-graded by an endomorphism $\gamma\in \Gamma_{C^{\infty}}(\IMM,\mathrm{End}(\IEE))$. The vector bundle $\mathscr{S}\to \IMM$ inherits a metric and a metric connection $\nabla$ from the Riemannian metric and the Levi-Civita connection on $\IMM$. Let 
$$
D: \Gamma_{C^{\infty}}(\IMM,\mathscr{S})\longrightarrow \Gamma_{C^{\infty}}(\IMM,\mathscr{S})
$$
denote the induced Dirac operator and let
\begin{align*}
&c:\Omega_{C^{\infty}}(M)\longrightarrow \Gamma_{C^{\infty}}(\IMM, \mathrm{End}(\mathscr{S})),\quad c(\alpha_1\wedge\cdots\wedge \alpha_p)\Psi:=\frac{1}{p!}\alpha_1\cdots \alpha_p \cdot \Psi,\\
&\quad \alpha_1,\dots,\alpha_p\in \Omega^1_{C^{\infty}}(M),\quad \Psi\in\Gamma_{C^{\infty}}(\IMM,\mathscr{S}),
\end{align*}
denote the natural extension of the (dual) Clifford multiplication 
$$
\Omega^1_{C^{\infty}}(M)\longrightarrow \Gamma_{C^{\infty}}(\IMM, \mathrm{End}(\mathscr{S})),\quad \alpha\longmapsto (\Psi\longmapsto \alpha\cdot \Psi)
$$
from $1$-forms to arbitrary differential forms. The operator $D$ (defined a priori on $\Gamma_{C^{\infty}}(\IMM,\mathscr{S})$) is essentially self-adjoint in $\Gamma_{L^2}(\IMM,\mathscr{S})$, and its unique self-adjoint realization will be denoted with the same symbol again. With $\T:=S^1$ let 
$$
 \Omega_\T(\IMM):= \Omega_{C^{\infty}}(\IMM \times \T)^{\T}
$$
denote the space of $\T$-invariant differential forms on $\IMM\times \T$. Each element $\alpha$ of $\Omega_\T(\IMM)$ can be uniquely written in the form $\alpha=\alpha'+\alpha''\Id t$ with $\Id t$ the volume form on $\T$. Define a complex linear space by
$$
 \mathsf{C}_\T(\IMM): = \bigoplus_{N=0}^\infty\Omega_\T(\IMM)\otimes\left(\Omega_\T(\IMM)^{\otimes N} /  (\C\cdot 1)  \right).
$$
Since, $\IMM$ is compact, $\mathrm{e}^{-t D^2}$ is trace class for all $t>0$. In this situation, the \emph{Chern Character} $\mathrm{Ch}_{\mathbb{T}}(\IMM)$ is a linear functional\footnote{In fact, $\mathrm{Ch}_{\mathbb{T}}(\IMM)$ extends continuously to a certain completion of $\mathsf{C}_\T(\IMM)$, but we shall not be concerned with this fact here.} 
$$
\mathrm{Ch}_{\mathbb{T}}(\IMM): \mathsf{C}_\T(\IMM)\longrightarrow \IC,
$$
that has been introduced in \cite{gl}. The formula for $\mathrm{Ch}_{\mathbb{T}}(\IMM)$ is given as follows: define
$$
F_{\T}:  \mathsf{C}_\T(\IMM)\longrightarrow \{\text{differential operators of order $\leq 2$ in $\mathscr{S}\to\IMM$}\}
$$
by
\begin{align*}
  %&F^{(0)}_{\T} = D^2,\\
  &F_{\T}(\alpha_0)= c(\Id \alpha_0^\prime) - [D, c(\alpha_0^\prime)] - c(\alpha_0^{\prime\prime})\\
  &F_{\T}(\alpha_0\otimes \alpha_1)= (-1)^{|\alpha_0^\prime|}\bigl(c({\alpha}_0^\prime \wedge  {\alpha}_1^\prime) - c({\alpha}_0^\prime)c({\alpha}_1^\prime)\bigr),\\
	&F_{\T}(\alpha_0\otimes\cdots\otimes \alpha_N)=0\quad\text{ for all $N\geq 3$.} 
\end{align*}
Above, $[D,c(\alpha)]$ denotes a $\mathbb{Z}_2$-graded commutator (where differential forms are $\IZ_2$-graded through even/odd form degrees). Explicitly, one has 
$$
[D,c(\alpha)]=Dc(\alpha)-(-1)^pc(\alpha)D,\quad\text{if $\alpha\in \Omega^p_{C^{\infty}}(\IMM)$.}
$$

For natural numbers $L\leq N $ denote with $\mathsf{P}_{L, N}$ all tuples $I=(I_1, \dots, I_L)$ of subsets of $\{1 \dots, N\}$ with 
$$
I_1 \cup \dots \cup I_L = \{1 \dots, N\}
$$
and with each element of $I_a$ smaller than each element of $I_b$ whenever $a < b$. Given 
\begin{itemize}
\item $\alpha_1\otimes\cdots\otimes \alpha_N\in\Omega_{\T}(\IMM)^{\otimes N}$,
\item $I=(I_1, \dots, I_L)\in \mathsf{P}_{L, N}$,
\item $1\leq a\leq L$, 
\end{itemize}
we set 
$$
\alpha_{I_a}:= \alpha_{i+1} \otimes \cdots \otimes \alpha_{i+l},\quad\text{ if $I_a = \{j \mid i < j \leq i+l\}$ for some $i, l$. }
$$
Then with $\mathrm{Str}(\bullet):=\mathrm{Tr}(\gamma\bullet)$ the $\mathbb{Z}_2$-graded trace on $\ILL(\Gamma_{L^2}(\IMM,\mathscr{S}))$, one has
\begin{align*}
&\mathrm{Ch}_{\mathbb{T}}(\IMM)(\alpha_0\otimes\cdots\otimes\alpha_N):=\sum_{L=1}^N (-1)^L \sum_{I \in \mathsf{P}_{L, N}} \int_{\{0\leq s_1\leq\cdots\leq s_L\leq 1\}} \mathrm{Str}\left(c(\alpha_0)\mathrm{e}^{-s_1 D^2} F_{\T}(\alpha_{I_1})\times \right.\\
&\left. \quad\quad\quad\quad\quad\quad\quad\quad\quad\quad\quad\times \mathrm{e}^{-(s_2-s_1)D^2}F_{\T}(\alpha_{I_2})\cdots \mathrm{e}^{-(s_L-s_{L-1})D^2} F_{\T}(\alpha_{I_L}) \mathrm{e}^{-(1-s_L)D^2}\right)  \Id s_1\cdots \Id s_L.
\end{align*}
By definition the $N=0$ part of the Chern character is given explicitly by
\begin{align}
\mathrm{Ch}_{\mathbb{T}}(\IMM)(\alpha_0)=\mathrm{Str}\left(c(\alpha_0')\mathrm{e}^{-D^2}\right),
\end{align}
and the $N=1$ part is given explicitly by 
\begin{align}
\mathrm{Ch}_{\mathbb{T}}(\IMM)(\alpha_0\otimes \alpha_1)=-\mathrm{Str}\left(\int^1_0c(\alpha_0')\mathrm{e}^{-sD^2}F_{\T}(\alpha_1) \mathrm{e}^{-(1-s)D^2} \Id s\right).
\end{align}
By the Lichnerowicz formula we have
\begin{align}\label{asay}
D^2=\nabla^{\dagger}\nabla+(1/4)\mathrm{scal},
\end{align}
so that the $N=0$ piece of $\mathrm{Ch}_{\mathbb{T}}(\IMM)$ is given by the probabilistic expression
$$
\mathrm{Ch}_{\mathbb{T}}(\IMM)(\alpha_0)=\int_M\mathrm{e}^{-tH}(x,x)\mathrm{Str}_{x}\left(c(\alpha_0')(x)\mathbb{E}^{x,x}_t\left[\mathrm{e}^{-(1/8)\int^t_0\mathrm{scal}(\bb^x_s)\Id s}\pa^{x}_{\nabla}(t)^{-1}\right]|_{t=2}\right)\Id\mu(x),
$$
with $\mathrm{Str}_{x}$ the $\IZ_2$-graded trace on $\mathrm{End}(\mathscr{S}_x)$. We are going to use Theorem \ref{main2} to deduce a probabilistic representation for the $N=1$ piece of $\mathrm{Ch}_{\mathbb{T}}(\IMM)$:

\begin{theorem}\label{main3} Let $\IMM$ be a compact even dimensional Riemannian spin manifold. Then for all $\alpha_0,\alpha_1\in \Omega_\T(\IMM)$ one has
\begin{tiny}
\begin{align*}
&\mathrm{Ch}_{\mathbb{T}}(\IMM)(\alpha_0\otimes\alpha_1)\\
&=\int_\IMM\mathrm{e}^{-tH}(x,x)\mathrm{Str}_x\left(c(\alpha_0')(x)\mathbb{E}^{x,x}_t\left[\mathrm{e}^{-(1/8)\int^t_0\mathrm{scal}(\bb^x_s)\Id s}\int^t_0 \transport^x_{\nabla}(s)^{-1} \Big( 2c(*\Id\bb_s^x\lrcorner \alpha_1')  -c(\alpha_1'')(\bb_s^x) \Id s\Big)\transport^x_{\nabla}(s)\transport^{x}_{\nabla}(t)^{-1}\right]|_{t=2}\right)\Id\mu(x).
\end{align*}
\end{tiny}
\end{theorem}

\begin{proof} Applying Theorem \ref{main2} with $V:=(1/8)\mathrm{scal}$, $\tilde{V}:=\gamma$ and $P:=F_\T(\alpha_1)$, and noting that by (\ref{asay}) one has $H^{\nabla}_V= D^2$, for all $x,y\in\IMM$, we immediately get  
\begin{tiny}
\begin{align*}
&\mathrm{Str}\left(\int^1_0\mathrm{e}^{-sD^2}F_{\T}(\alpha_1) \mathrm{e}^{-(1-s)D^2} \Id s\right)=\\
&\int_\IMM\mathrm{e}^{-tH}(x,x)\mathbb{E}^{x,y}_t\left[\mathrm{e}^{-(1/8)\int^t_0\mathrm{scal}(\bb^x_s)\Id s}\int^t_0 \transport^x_{\nabla}(s)^{-1} \big( \sigma_1(F(\alpha_1))^{\flat}(\Id\bb_s^x)+ F(\alpha_1)_{\nabla}(\bb_s^x) \Id s\big)\transport^x_{\nabla}(s)\transport^{x}_{\nabla}(t)^{-1}\right]|_{t=2}\Id\mu(x).
\end{align*}
\end{tiny}
With the product
\[
\star: \Gamma_{C^{\infty}}(\IMM,T\IMM\otimes \mathscr{S}) \otimes\Omega_{C^{\infty}}(\IMM)\longrightarrow  \Gamma_{C^{\infty}}(\IMM, \mathscr{S}) ,\quad 	(X\otimes \varphi)\star \alpha := c(X\lrcorner \alpha)\varphi, 
\]
where $X\lrcorner \alpha$ denotes the contraction of the form $\alpha$ by the vector field $X$, we are going to prove in a moment the formula
\begin{align}\label{ssppaa}
[D,c(\alpha)]\varphi = c((\Id+\Id^{\dagger})\alpha)\varphi-2 (\nabla\varphi)^{\sharp\otimes\mathrm{Id}}\star\alpha, \quad\alpha\in\Omega_{C^{\infty}}(\IMM),\quad \varphi\in\Gamma_{C^{\infty}}(\IMM,\mathscr{S}).
\end{align}
Given this identity, we find 
$$
\sigma_1(F_{\T}(\alpha_1))^{\flat}(X)= 2c(X\lrcorner \alpha_1' )\quad\text{for all vector fields $X$ on $\IMM$,}
$$
 and furthermore 
$$
F_{\T}(\alpha_1)_{\nabla}= -c(\Id^{\dagger}\alpha_1')-c(\alpha_1''),
$$
so that the above is
\begin{tiny}
$$
=\int_\IMM\mathrm{e}^{-tH}(x,x)\mathbb{E}^{x,x}_t\left[\mathrm{e}^{-(1/8)\int^t_0\mathrm{scal}(\bb^x_s)\Id s}\int^t_0 \transport^x_{\nabla}(s)^{-1} \Big( 2c(\Id\bb_s^x\lrcorner \alpha_1')  -c(\Id^{\dagger}\alpha_1')(\bb_s^x)-c(\alpha_1'')(\bb_s^x) \Id s\Big)\transport^x_{\nabla}(s)\transport^{x}_{\nabla}(t)^{-1}\right]|_{t=2}\Id\mu(x).
$$ 
\end{tiny}
Using the It\^{o}-to-Stratonovic rule
$$
c(\Id \bb^x_s  \lrcorner \alpha')= c(  *\Id \bb^x_s \lrcorner\alpha')+\frac{1}{2}c(\Id^{\dagger}\alpha')(\bb_s^x)\Id s,
$$
we arrive at
\begin{tiny}
\begin{align*}
&\mathrm{Str}\left(\int^1_0\mathrm{e}^{-D^2}F_{\T}(\alpha_1) \mathrm{e}^{-(1-s)D^2} \Id s \right)\\
&=\int_\IMM\mathrm{e}^{-tH}(x,x)\mathbb{E}^{x,y}_t\left[\mathrm{e}^{-(1/8)\int^t_0\mathrm{scal}(\bb^x_s)\Id s}\int^t_0 \transport^x_{\nabla}(s)^{-1} \Big( 2c(*\Id\bb_s^x\lrcorner \alpha_1')  -c(\alpha_1'')(\bb_s^x) \Id s\Big)\transport^x_{\nabla}(s)\transport^{x}_{\nabla}(t)^{-1}\right]|_{t=2}\Id\mu(x),
\end{align*}
\end{tiny}
which is the claimed formula. \\
It remains to prove (\ref{ssppaa}). To this end, denote with $\Cl(\IMM)\to \IMM$ the Clifford bundle and with 
$$
\tilde{}:\Omega_{C^{\infty}}(\IMM)\longrightarrow \Gamma_{C^{\infty}}(\IMM, \Cl(\IMM))
$$
the natural isomorphism. Then we have 
$$
\widetilde{(\dd+\Id^{\dagger})\alpha}=D^{\Cl(M)}\tilde{\alpha}
$$ 
(cf. \cite{LM89}, Chapter~II, Thm.~5.12), with $D^{\Cl(M)}$ the natural Dirac operator on $\Cl(\IMM)\to \IMM$. Assume now $\alpha\in\Omega^p(\IMM)$ and pick a local orthonormal frame $(e_1,\ldots,e_m)$. Write $\alpha = \sum_{I}\alpha_I \mathrm{e}^*_{i_1}\wedge\ldots\wedge e_{i_p}^*$ with some increasingly ordered multi-index $I=(i_1,\ldots,i_p)$. One has
	\begin{align}
	&[D,c(\alpha)]\varphi = Dc(\alpha)\varphi - (-1)^p c(\alpha)\varphi \\
	&= \sum_{j=1}^{n}\sum_{I}\left(e_j\cdot\nabla_{e_j}(\alpha_Ie_{i_1}\cdots e_{i_p}\cdot\varphi) + (-1)^{p+1}\alpha_Ie_{i_1}\cdots e_{i_p}\cdot e_j\cdot\nabla_{e_j}\varphi \right)\nonumber\\
	&= \sum_{j=1}^{n}\sum_{I}\left(e_j\cdot\nabla^{\Cl(\IMM)}_{e_j}(\alpha_I e_{i_1}\cdots e_{i_p})\cdot\varphi + e_j\cdot\alpha_I e_{i_1}\cdots e_{i_p}\nabla_{e_j}\varphi + (-1)^{p+1}\alpha_Ie_{i_1}\cdots e_{i_p}\cdot e_j\cdot\nabla_{e_j}\varphi\right)\nonumber\\
	&= \sum_{j=1}^{n}\sum_{I}\left(e_j\cdot\nabla^{\Cl(\IMM)}_{e_j}(\alpha_I e_{i_1}\cdots e_{i_p})\cdot\varphi + \alpha_I(e_j\cdot e_{i_1}\cdots e_{i_p} + (-1)^{p+1}e_{i_1}\cdots e_{i_p}\cdot e_j	)\cdot \nabla_{e_j}\varphi \right)\nonumber\\
	&= (D^{\Cl(\IMM)}\tilde{\alpha})\cdot \varphi + \sum_{j=1}^{n}\sum_{I}\left(\alpha_I(e_j\cdot e_{i_1}\cdots e_{i_p} + (-1)^{p+1}e_{i_1}\cdots e_{i_p}\cdot e_j	)\cdot \nabla_{e_j}\varphi \right)\,.\label{eqn:sk0}
	\end{align}
	Fix now $I$ and $j$. In case $j\neq i_k$ for all $k=1,\ldots,p$, one has
	\begin{align*}
		e_j\cdot e_{i_1}\cdots e_{i_p} & = (-1)^p e_{i_1}\cdots e_{i_p}\cdot e_j \,.
	\end{align*}
	In case $j=i_k$ for some $1\leq k\leq p$, one has
	\begin{align*}
		e_j\cdot e_{i_1}\cdots e_{i_p} & = e_{i_k}\cdot e_{i_1}\cdots e_{i_p} = (-1)^{k-1}e_{i_1}\cdots e_{i_k}\cdot e_{i_k}\cdots e_{i_p} = (-1)^k e_{i_1}\cdots \widehat{e_{i_k}}\cdots e_{i_p}
	\end{align*}
	and
	\begin{align*}
		(-1)^{p+1}e_{i_1}\cdots e_{i_p}\cdot e_j &= (-1)^{p+1}e_{i_1}\cdots e_{i_p}\cdot e_{i_k} = (-1)^{p+1+p-k}e_{i_1}\cdots e_{i_k}\cdot e_{i_k}\cdots e_{i_p} \\
		&= (-1)^{2p+2-k}e_{i_1}\cdots \widehat{e_{i_k}}\cdots e_{i_p}=(-1)^{k}e_{i_1}\cdots \widehat{e_{i_k}}\cdots e_{i_p}\,.
	\end{align*}
	So the RHS of \eqref{eqn:sk0} equals
	\begin{align}
		c((\dd+\Id^{\dagger})\alpha)\varphi - 2\sum_I\sum_{k=1}^{p}(-1)^{k-1}\alpha_Ie_{i_1}\cdots \widehat{e_{i_k}}\cdots e_{i_p} \cdot \nabla_{e_{i_k}}\varphi\,.\label{eqn:sk1}
	\end{align}
	Assume again $I$ and $j$ are fixed and that $j=i_k$ for some $k$. Then by the definition of the product $\star$,
	\begin{align*}
&	(e_j\otimes \nabla_{e_{j}}\varphi)\star \alpha_I e_{i_1}^*\wedge\ldots\wedge e_{i_p}^* &= c(e_j \lrcorner \alpha_I e_{i_1}^*\wedge\ldots\wedge e_{i_p}^*)\nabla_{e_j}\varphi \\
	&= c((-1)^{k-1}\alpha_I e_{i_1}^*\wedge\ldots  \wedge \widehat{e_{i_k}}\wedge\ldots\wedge e_{i_p}^*)\nabla_{e_{i_k}}\varphi\\
	&=(-1)^{k-1}\alpha_Ie_{i_1}\cdots \widehat{e_{i_k}}\cdots e_{i_p} \cdot \nabla_{e_{i_k}}.
	\end{align*} 
	As one has $(\nabla\varphi)^{\sharp\otimes\Id} = \sum_{j=1}^{n}e_j\otimes \nabla_{e_j}\varphi$, \eqref{eqn:sk1} equals
	\[
		c((\dd+\Id^{\dagger})\alpha)\varphi - 2\sum_I\sum_{j=1}^{n} (e_j\otimes\nabla_{e_{j}}\varphi)\star \alpha_I\mathrm{e}^*_{i_1}\wedge\cdots\wedge  \mathrm{e}^*_{i_p} = c((\dd+\Id^{\dagger})\alpha)\varphi-2(\nabla\varphi)^{\sharp\otimes\Id}\star\alpha  \,,\label{eqn:sk2}
	\]
	completing the proof.
\end{proof}

\appendix

\section{Facts on sectorial forms and operators}\label{saab}

In this appendix, we have collected some definitions and facts on sectorial forms and operators, following the presentation from section VI in \cite{kato}. \\
A densely defined sesqui-linear form $h$ in a Hilbert space $\IHH$ is called \emph{sectorial}, if there exist numbers $\beta\in [0,\pi/2), \gamma\in\IR$ such that
$$
 \{h(\Psi,\Psi): \Psi\in \dom(h), \left\|\Psi\right\|=1\}\subset \{z\in\IC:|\mathrm{arg}(z-\gamma)|\leq \beta\}.
$$
Above, $\gamma$ is called a \emph{vertex} of $h$ and $\beta$ an \emph{angle} of $h$.\vspace{1mm}

A sectorial form $h$ in $\IHH$ is called \emph{closed} if for all $\Psi\in\IHH$ which admit a sequence $(\Psi_n)\subset \dom(h)$ with 
$$
\left\|\Psi_n-\Psi\right\|\to 0,\quad h(\psi_n-\psi_l, \psi_n-\psi_l)\to 0 \quad\text{as $n,l\to\infty$},
$$
and $h$ is called \emph{closable} if it has a closed extension; in this case $h$ has a smallest closed extension $\overline{h}$, called the \emph{closure} of $h$. Sums of sectorial forms are sectorial, and sums of closed forms are closed (on their natural domain of definition; Theorem 131 p. 319 in \cite{kato}). \vspace{1mm}

A densely defined operator $S$ in $\IHH$ is called \emph{sectorial}, if the form $h_S$ given by $\dom(h_S)=\dom(S)$ and $h_S(\Psi_1,\Psi_2)=\left\langle S\Psi_1,\Psi_2\right\rangle$ is sectorial. If a form $h$ in $\IHH$ is induced by a sectorial operator $S$ in $\IHH$, in the sense that $h=h_S$, then $h$ is closable (Theorem 1.27 p. 318 in \cite{kato}). 

\vspace{1.2mm}

\begin{theorem}\label{klmn} If $h$ is sectorial and the form $h'$ satisfies $\dom(h)\subset \dom(h')$ and admits constants $a\in [0,\infty)$, $b\in [0,1)$ such that
$$
|h'(\Psi,\Psi)|\leq a\left\|\Psi\right\|^2+ b |h(\Psi,\Psi)|\quad\text{ for all $\Psi\in\dom(h)$,}
$$
then the form $h+h'$ is 
\begin{itemize}
\item sectorial,
\item closed if and only if $h$ is closed,
\item closable if and only if $h$ is closable; and then $\dom(\overline{h+h'})=\dom(\overline{h})$.
\end{itemize}

\end{theorem}

\begin{proof} This is Theorem 1.33 p. 320 in \cite{kato}.
\end{proof}

Given $\beta\in (0,\pi/2]$ set
$$
 \Sigma_{\beta}=\{r\mathrm{e}^{\sqrt{-1}\alpha}:r> 0, \alpha\in (-\beta,\beta)\}
$$
and
$$
\Sigma_{0,\beta}:= \Sigma_{\beta}\cup\{0\}=\{r\mathrm{e}^{\sqrt{-1}\alpha}:r\geq 0, \alpha\in (-\beta,\beta)\}.
$$

A family of bounded operators $(T_z)_{z\in \Sigma_{0,\beta}}$ in $\IHH$, with some $\beta\in (0,\pi/2]$, is called a \emph{holomorphic semigroup}, if
\begin{itemize}
\item $z\mapsto T_z$ is holomorphic\footnote{Here, weak/strong/norm holomorphy are equivalent by the uniform boundedness principle.} in $z\in \Sigma_{\beta}$,
\item $T_{z+z'}=T_{z}T_{z'}$ for all $z,z'\in \Sigma_{0,\beta}$,
\item $z\mapsto T_z$ is strongly continuous in $z=0$ and $T(0)=1$.
\end{itemize}

It follows that the restriction of $T$ to $[0,\infty)$ is a strongly continuous semigroup, and if $S$ is the generator of this semigroup, then for every $\Psi_0\in \IHH$, the function 
$$
[0,\infty)\ni t\longmapsto T(t)\Psi_0\in \IHH
$$
is the uniquely determined strongly continuous function $\Psi:[0,\infty)\to \Gamma_{L^2}(\IMM,\IEE)$ which is strongly differentiable on $(0,\infty)$ taking values in $\dom(S)$ thereon, such that
$$
(\Id /\Id t) \Psi(t)=S\Psi(t),\quad t>0,\quad \Psi(0)=\Psi_0.
$$
Thus, one essential property of holomorphic semigroups is that the above initial value problem has a unique solution for every initial value in $\IHH$, rather than just for initial values in the domain of the generator (cf. Remark 1.22 on p. 492 in \cite{kato}).\vspace{1mm}

Finally, there is the following representation theorem:

\begin{theorem}\label{gene} For every closed sectorial form $h$ in $\IHH$ there exists a unique densely defined, closed, and sectorial operator $S$ in $\IHH$ such that $\dom(S)\subset \dom(h)$ and 
\begin{align}\label{klq}
h(\Psi_1,\Psi_2)= \left\langle S\Psi_1,\Psi_2\right\rangle\quad\text{for all $\Psi_1\in \dom(S),\Psi_2\in \dom (h)$.}
\end{align}
Moreover, $-S$ generates a holomorphic semigroup in $\IHH$, to be denoted with $ z\mapsto \mathrm{e}^{-zS}$. 
\end{theorem}

\begin{proof} The existence of a densely defined, closed, and sectorial $S$ satisfying (\ref{klq}) is the statement of Theorem 2.1 on p. 322 in \cite{kato}. In fact, it is stated there that $S$ is actually $m$-sectorial, which by Theorem 1.24 on p. 492 in \cite{kato} implies that $-S$ generates a holomorphic semigroup, as for some $r\in\IR$, the form induced by $S+r$ has a vertex $0$  (see also Theorem 1.14 in (\cite{arendt})).
\end{proof}

\section{Some formulae for second order differential operators}\label{bochner}

Given a Riemannian manifold $\IMM$ with its volume measure $\mu$, let $\IEE\to \IMM$ be a vector bundle which is equipped with a covariant derivative $\nabla$.

\subsection{Nondivergence form}

Suppose we are given a second order linear differential operator $L$ acting on the sections of $\IEE\to \IMM$ via
\begin{align*}
	L = -\mathrm{tr}\nabla^2+ A \nabla + V
\end{align*}
with $A\in\Gamma_{C^\infty}(\IMM,\Hom(T^*\IMM\otimes \IEE,\IEE))$ and $V\in \Gamma_{C^\infty}(\IMM,\End(\IEE))$.

We introduce the notation
\[
A_X := A(X^\flat\otimes (\cdot) ) 
\]
for all smooth vector fields $X$ on $\IMM$, and we define a new covariant derivative on $\IEE\to \IMM$ by
\begin{align*}
	\nabla^A_X:=\nabla_X - \tfrac 12 A_X\,.
\end{align*}
Then, with a local orthonormal frame $(e_1,\ldots,e_n)$ for $T\IMM\to \IMM$, we calculate straightforwardly for every $\varphi\in \Gamma_{C^\infty}(\IMM,\IEE)$,
\begin{align*}
	-\mathrm{tr}(\nabla^A)^2\varphi &= -\sum_{i=1}^n \left(\nabla^A_{e_i}\nabla^A_{e_i}\varphi - \nabla^A_{\nabla_{e_i}e_i}\varphi\right) \\&= -\sum_{i=1}^n \left(\nabla_{e_i}-\tfrac 12A_{e_i}\right)\left(\nabla_{e_i}-\tfrac 12A_{e_i}\right)\varphi + \sum_{i=1}^n\left({\nabla}_{\nabla_{e_i}e_i}-\tfrac 12A_{\nabla_{e_i}e_i}\right)\varphi\\
	&= -\mathrm{tr}\nabla^2\varphi + \sum_{i=1}^n\left(\tfrac 12 A_{e_i}\nabla_{e_i}\varphi + \tfrac 12\nabla_{e_i}(A_{e_i}\varphi) - \tfrac 14 A_{e_i}^2\varphi -\tfrac 12 A_{\nabla_{e_i}e_i}\varphi\right)\\
	&= -\mathrm{tr}\nabla^2\varphi + \sum_{i=1}^n\left(\tfrac 12 A_{e_i}\nabla_{e_i}\varphi + \tfrac 12\nabla_{e_i}(A_{e_i}) \varphi + \tfrac 12A_{e_i}\nabla_{e_i}\varphi - \tfrac 14 A_{e_i}^2\varphi -\tfrac 12 A_{\nabla_{e_i}e_i}\varphi\right)\\
	&= -\mathrm{tr}\nabla^2\varphi + A\nabla\varphi + \sum_{i=1}^n\left(\tfrac 12\nabla_{e_i}(A_{e_i}) \varphi - \tfrac 14 A_{e_i}^2\varphi -\tfrac 12 A_{\nabla_{e_i}e_i}\varphi\right)\,,
\end{align*}
so that
\begin{align*}
	L = -\mathrm{tr}\ (\nabla^A)^2 + V^{\nabla,A},
\end{align*}
where $V^{\nabla,A}\in \Gamma_{C^\infty}(\IMM,\End(\IEE))$ is given by 
$$
V^{\nabla,A}:=V-\sum_{i=1}^n\left(\tfrac 12\nabla_{e_i}(A_{e_i})  - \tfrac 14 A_{e_i}^2 -\tfrac 12 A_{\nabla_{e_i}e_i}\right).
$$

\subsection{Divergence form}

We now assume additionally that our vector bundle $\IEE\to\IMM$ is endowed with a bundle metric $h(\cdot,\cdot)$. We do not, however, assume any compatibility between $\nabla$ and $h$, i.e.\ in general we do not have
\[
Xh(\varphi,\psi) = h(\nabla_X\varphi,\psi) + h(\varphi,\nabla_X\psi)\,,
\]
for all $\varphi, \psi\in \Gamma_{C^\infty}(\IMM,\IEE)$ and all smooth vector fields $X$ on $\IMM$.

Let us first prove that the formal adjoint $\nabla^\dagger$ of $\nabla$ is given by
\begin{align*}
	\nabla^\dagger(X^\flat\otimes\varphi) = -\nabla_X\varphi - \mathrm{div} X\cdot\varphi - ((\nabla_Xh )(\cdot,\varphi))^\sharp\,,
\end{align*}
for all $\varphi$ and $X$ as above. Indeed, using $X(f) + f\mathrm{div} X = \mathrm{div}(fX)$ for all smooth functions $f$ on $\IMM$, we have for all smooth compactly supported sections $\psi$ of $\IEE\to\IMM$,
\begin{align*}
	\int_\IMM\! h(\varphi,\nabla^{\dagger}(X^\flat\otimes\psi))\dd\mu &= \int_\IMM\!h\left(\varphi,-\nabla_X\psi - \div X\cdot\psi - ((\nabla_Xh )(\cdot,\psi))^\sharp\right)\dd\mu\\
	&= \int_\IMM\!\left(-h(\varphi,\nabla_X\psi) - h(\varphi,\psi)\cdot\div X - (\nabla_Xh)(\varphi,\psi) \right)\dd\mu\\
	&= \int_\IMM\!(-h(\varphi,\nabla_X\psi) - h(\varphi,\psi)\cdot\div X - Xh(\varphi,\psi)+h(\nabla_X\varphi,\psi)\\&\qquad +h(\varphi,\nabla_X\psi) )\dd\mu\\
	&=\int_\IMM\!\left(h(\nabla_X\varphi,\psi)-\div\left(h(\varphi,\psi)X\right)\right)\dd\mu\\
	&=\int_\IMM\!h(\nabla_X\varphi,\psi)\dd\mu\\
	&=\int_\IMM\!h(\nabla\varphi,X^\flat\otimes\psi)\dd\mu\,.
\end{align*}

Let now a second order linear differential operator $L$ be given by
\begin{align*}
	L = \nabla^{\dagger}\nabla + A\nabla +V
\end{align*}
with $A,V$ as in the previous section. Note that now the second order part of $L$ is in divergence form, in contrast to the previous section. We are going to carry out a calculation similar to the one in the previous section to know exactly for which $A$'s the operator 
\begin{align}\label{order}
L':=L - (\nabla^A)^{\dagger}\nabla^A\quad\text{is of zeroth order.}
\end{align}

First of all, one has
\begin{align*}
	(\nabla^A_X h)(\varphi,\psi) &= Xh(\varphi,\psi)-h(\nabla^A_X\varphi,\psi)-h(\varphi,\nabla^A_X\psi)\\
	&= Xh(\varphi,\psi) - h(\nabla_X\varphi,\psi) - h(\varphi,\nabla_X\psi) + \tfrac 12h(A_X\varphi,\psi) + \tfrac 12 h(\varphi,A_X\psi)\\
	&= (\nabla_Xh)(\varphi,\psi) + h(\varphi,\tfrac 12 (A_X+A_X^{\dagger})\psi)\,,
\end{align*}
which, in turn, implies
\begin{align*}
	(\nabla^A)^{\dagger}(X^\flat\otimes\varphi) = \nabla^{\dagger}(X^\flat\otimes\varphi)-\tfrac 12 A_X^{\dagger}\varphi\,.
\end{align*}

A calculation analogous to the one in the previous section then yields
\begin{align*}
	&(\nabla^A)^{\dagger}\nabla^A\varphi \\
	&= \nabla^{\dagger}\nabla\varphi + \sum_{i=1}^n\tfrac 12 (A_{e_i}-A_{e_i}^{\dagger})\nabla_{e_i}\varphi  +\tfrac12\sum_{i=1}^n\big(\nabla_{e_i}(A_{e_i})\varphi +\div e_i\cdot A_{e_i}\varphi + (\nabla_{e_i}h)(\cdot,A_{e_i}\varphi)^\sharp \big)\\
	&\quad + \sum_{i=1}^n\tfrac12 A_{e_i}^{\dagger}A_{e_i}\varphi\\
	&= \nabla^{\dagger}\nabla\varphi + \tfrac12(A-A^{\dagger})\nabla\varphi +\tfrac12\sum_{i=1}^n\big(\nabla_{e_i}(A_{e_i})\varphi +\div e_i\cdot A_{e_i}\varphi + (\nabla_{e_i}h)(\cdot,A_{e_i}\varphi)^\sharp + \tfrac12 A_{e_i}^{\dagger}A_{e_i}\big)\varphi,
\end{align*}
where we suggestively wrote $A^{\dagger}$ for the section defined by $A^{\dagger}(X^\flat\otimes\varphi)=A_X^{\dagger}\varphi$. Note that $A-A^{\dagger}$ is skewsymmetric. 

The above calculation shows that, if $A\neq 0$, then one has (\ref{order}) if and only if $A_X$ is skewsymmetric for all $X$.

\end{document}